\documentclass{siamart1116}
\usepackage{indentfirst}
\usepackage{graphicx}
\usepackage{epstopdf}
\usepackage{epsfig}
\usepackage{overpic}
\usepackage{overcite}
\usepackage{array}
\usepackage{color}
\usepackage{multirow}
\usepackage{float}
\usepackage{subfigure} 
\usepackage{amsmath,amssymb}
\usepackage[mathscr]{euscript}
\usepackage{latexsym,bm}
\usepackage{booktabs}
\usepackage{bbm}
\usepackage{diagbox}
 \usepackage[all]{xy}
 \usepackage{color}
\usepackage{diagbox}
\usepackage{algorithm}  
\usepackage{algpseudocode}  
\usepackage{hyperref}
\usepackage{graphicx}
\usepackage{mathtools}
\def\dif{\mathrm{d}}
\def\mi{\mathrm{i}}
\def\me{\mathrm{e}}

\makeatletter
\@addtoreset{equation}{section}
\makeatother
\renewcommand{\theequation}{\arabic{section}.\arabic{equation}}

\begin{document}
\bibliographystyle{unsrt}
\title{Nonlocalization of singular potentials in quantum dynamics}
\author{Sihong Shao\footnotemark[1],
Lili Su\footnotemark[2]
}
\renewcommand{\thefootnote}{\fnsymbol{footnote}}
\footnotetext[1]{CAPT, LMAM and School of Mathematical Sciences, Peking University, Beijing 100871, China. Email: {\tt sihong@math.pku.edu.cn}.}
\footnotetext[2]{School of Mathematical Sciences, Peking University, Beijing 100871, China. Email: {\tt sull@stu.pku.edu.cn}.}
\maketitle

\begin{abstract}
Nonlocal modeling has drawn more and more attention and becomes steadily more powerful in scientific computing. 
In this paper, we demonstrate the superiority of a first-principle nonlocal model --- Wigner function --- in treating singular potentials which are often used to model the interaction between point charges in quantum science. 
The nonlocal nature of the Wigner equation is fully exploited to convert the singular potential into the Wigner kernel with weak or even no singularity, and thus highly accurate numerical approximations are achievable,
which are hardly designed when the singular potential is taken into account in the local Schr\"odinger equation.
The Dirac delta function, the logarithmic, and the inverse power potentials are considered. 
Numerically converged Wigner functions under all these singular potentials are obtained with an operator splitting spectral method, and display many interesting quantum behaviors as well.

\vspace*{4mm}

\end{abstract}

\newsavebox{\tablebox}
\setcounter{tocdepth}{3}

\section{Background and motivation}
\label{sec:intr}

It has been shown that the point charge description of electrons usually agrees well with the experimental results \cite{visscher1997dirac}, where the interaction between them is dominated by the Coulomb potential --- a typical singular potential in quantum science \cite{roy2005studies,case1950singular, perelomov1970fall, meetz1964singular, frank1971singular}. 
Such Coulomb interaction has found various applications in physics \cite{roy2005studies, keraani2005wigner} and chemistry \cite{visscher1997dirac, cinal2020highly, wei1999discrete}. Apart from that, there exist some other singular potentials to describe the interactions arising from scattering problems \cite{esposito1998scattering},
the short-range interactions in condensed matter \cite{tan2008energetics, gusson2018dirac},
Dirac monopole in the magnetic field \cite{ray2014observation} and etc.
The logarithmic potential is also adopted to measure the entropy density in the study of two-phase flow \cite{gal2017nonlocal}.

\begin{figure}
\centering 
\subfigure[The Dirac delta function potential \eqref{delta-pot}.]
{\includegraphics[width=0.495\textwidth,height=0.4\textwidth]{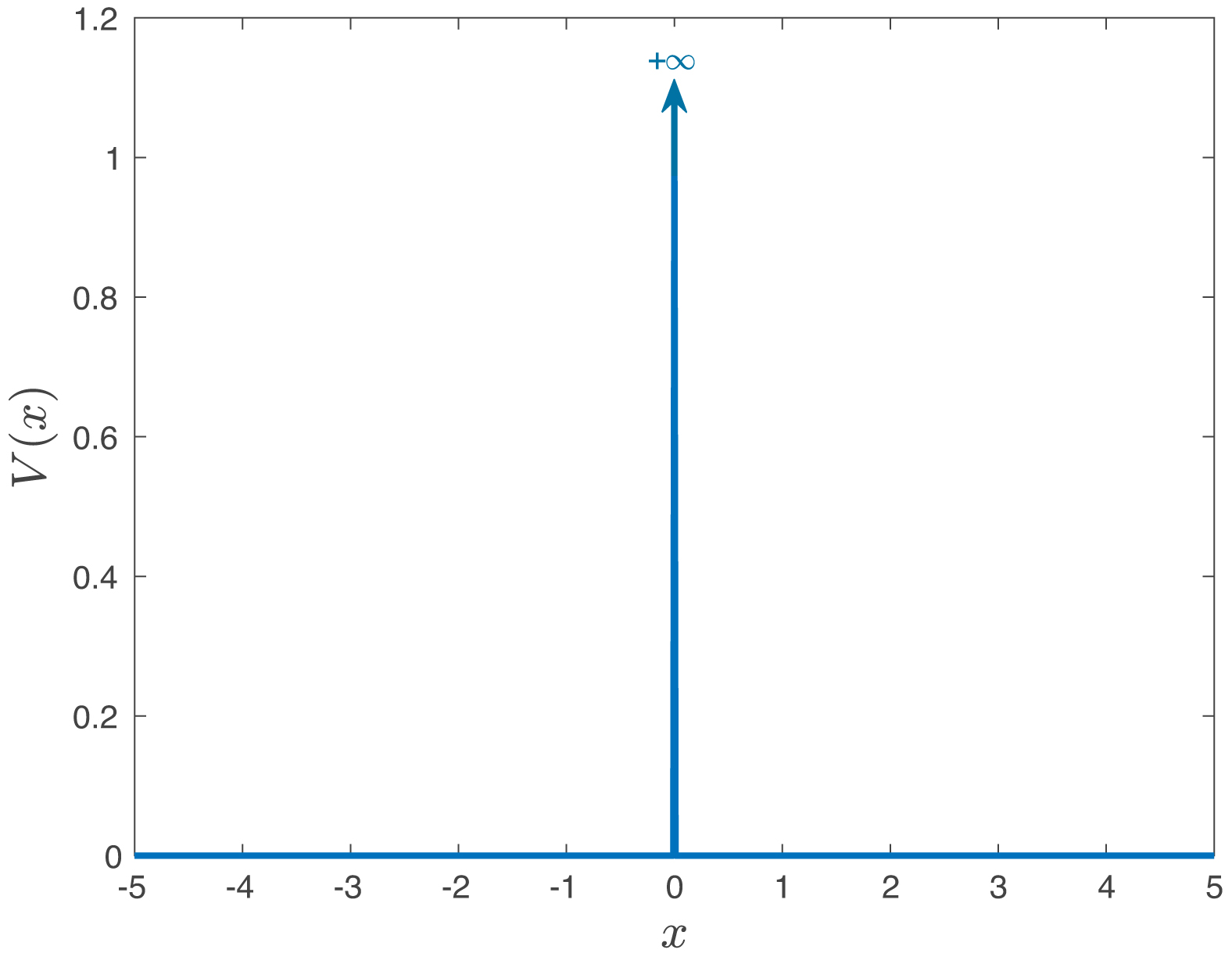}}
\subfigure[The Wigner kernel \eqref{Vw_delta}.]
{\includegraphics[width=0.495\textwidth,height=0.4\textwidth]{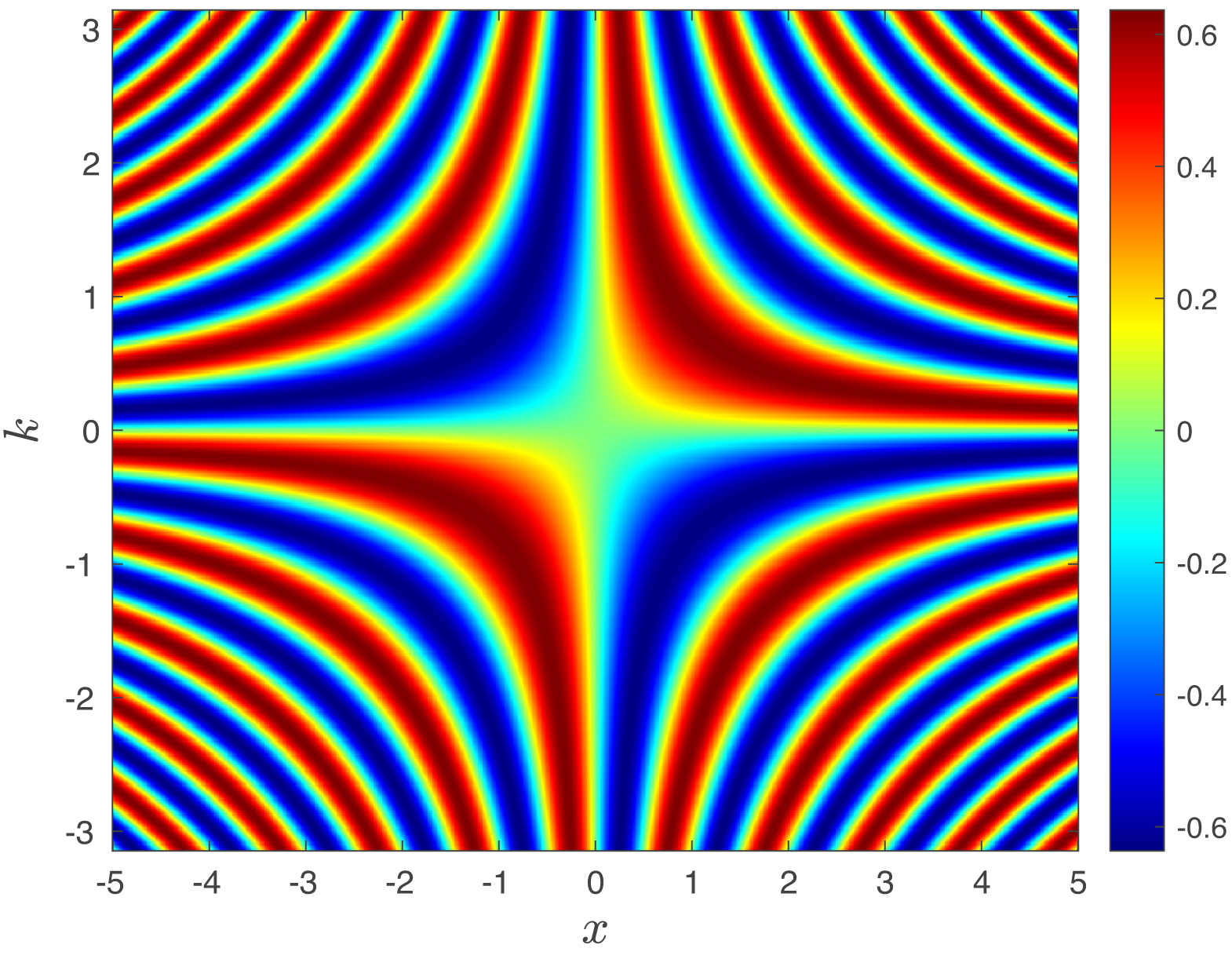}}
\caption[Vw\_delta]{\small The Dirac delta function potential and its Wigner kernel with power $H=1$.  It can be  intuitively seen that the singular potential is transformed into a non-singular Wigner kernel.} 
\label{fig:Vw_delta} 
\end{figure}

Directly plugging a singular potential into the Schr\"odinger equation 
\begin{equation}
\label{eq:SEQ_delta}
 \mi \hbar \frac{\partial}{\partial t} \psi(x,t) = -\frac{\hbar^2}{2m} \nabla_x^2 \psi(x,t) + V(x) \psi(x,t),
\end{equation}
and then seeking the numerical solutions runs into a problematic situation,
where $\psi(x,t)$ gives the wavefunction, $m$ and  $\hbar$  signify the mass of the particle and the reduced Planck constant, respectively.
Let's take the Dirac delta function potential as an example: 
\begin{equation}
\label{delta-pot}
V(x) = H \delta(x)
\end{equation}
with a power of $H$ (the influence of potential) and the Dirac delta function (see Fig.~\ref{fig:Vw_delta}): 
\begin{equation}
\label{delta}
\delta(x) = 
\begin{cases} 
+\infty & x=0 \\
0        & x\neq 0
\end{cases}, 
\quad
\int_{\mathbb{R}} \delta(x) \dif x = 1. 
\end{equation}
This Dirac delta function potential, which diverges at $x=0$,  is often adopted to model an infinite well and barrier \cite{belloni2014infinite}.
Obviously, there is no way for the finite difference method to find a suitable approximation to the function \eqref{delta} and then to the equation \eqref{eq:SEQ_delta} equipped with the singular potential \eqref{delta-pot}. 
Recourses to the Galerkin method inevitably sacrifices the accuracy or convergence order, which has been already pointed out by \cite{araya2006posteriori, scott1973finite} in solving the elliptical boundary value problem with the Dirac delta function source $-\Delta u(x) = \delta(x)$.

In this paper, we adopt a nonlocalization approach based on the integral formulation 
to deal with the singular potential situation.  
Specifically, we turns to the Wigner function \cite{PhysRev.40.749}
\begin{equation}
\label{WF}
 f(x,k,t) = \int_{\mathbb{R}} \psi(x+\frac{y}{2},t)\, \psi^\dagger(x-\frac{y}{2}, t)\, \exp(-\mi k y) ~\dif y,
\end{equation}
and its governing equation
\begin{equation}
\label{eq:WEQ}
\frac{\partial }{\partial t }f(  x,  k,t)
+ \frac{ \hbar   k}{m} \nabla_x f(  x,  k,t) = \Theta_{V}[f]( x, k, t),
\end{equation}
both of which are defined in phase space $(x,k)$ with
$x$ being the position and $k$ the wavenumber.
Starting from the definition \eqref{WF} where the Wigner function is calculated from the density matrix $\psi(x,t) \psi^\dagger(x, t)$ 
by changing to center-of-mass coordinates followed by a Fourier transform, the Wigner equation \eqref{eq:WEQ} can be derived from the Schr\"odinger equation \eqref{eq:SEQ_delta} in a straightforward manner.
The nonlocalization of singular potentials is embodied in the pseudo-differential operator
\begin{align}
    \Theta_{V}[f](  x,   k, t) & = \int_{\mathbb{R}} V_w(  x,  k- {k^\prime}) f(x,  k^\prime, t) ~\dif  k^\prime, \label{Th} \\
V_w(  x,  k) & =\frac{1}{2\pi \mi \hbar} \int_{\mathbb{R}} \exp(-\mi  k y) \,\left[ V(  x+\frac{   y}{2})-V(  x-\frac{   y}{2}) \right] \dif y, \label{Vw}
\end{align}
and all the information of potential $V(x)$ is contained in the Wigner kernel  $V_w(x, k)$.
Substituting the Dirac delta function potential \eqref{delta-pot} into Eq.\eqref{Vw} leads to 
\begin{equation}
\label{Vw_delta}
V_w(x, k) = \frac{2 H}{\pi \hbar} \sin(2xk),
\end{equation}
the plot of which is displayed in Fig.~\ref{fig:Vw_delta}. It can be readily observed there that the Wigner kernel  $V_w(x,  k)$ is no longer singular and thus we have a chance to seek highly accurate numerical solutions to the Wigner equation \eqref{eq:WEQ} with singular potentials. 
That is, the point singularity in $V(x)$ is distributed over the whole space with the nonlocal action of pseudo-differential operator, thereby alleviating or even eliminating the singularity. 
After obtaining the Wigner function \eqref{WF}, the average of a quantum operator $\hat{A}$ can be expressed as 
\begin{equation}
\label{average_A_wf}
\langle \hat{A} \rangle_t= \iint_{\mathbb{R} \times \mathbb{R}} A(x, k) f(x, k, t) ~\dif x \dif k, 
\end{equation}
where $A(x, k)$ gives the corresponding classical function in phase space.
In other words, the Wigner function formulation is fully equivalent to the wavefunction formulation for quantum mechanics \cite{curtright2013concise}.  Generally speaking, nonlocal models may offer more explanations for phenomena that involve possible singularities including the interaction with singular potentials and occurring at a distance \cite{nonlocal2020}.

By exploiting the intrinsic nonlocal nature of the Wigner function approach,
we are able to obtain highly accurate numerical approximations to observable quantities in quantum dynamics with singular potentials 
with the help of spectral methods and operator splitting techniques.  
For demonstration purposes, this work focuses on the singular potentials the Wigner kernels of which have analytical forms,
like the Dirac delta function potential. Otherwise, other extra numerical techniques should be adopted, for example, the truncated kernel method \cite{greengard2018anisotropic, vico2016fast}.
It should be noted that there exist few high precision numerical simulations of the Wigner equation under singular potentials except for a recent attempt to numerically solve the Wigner-Coulomb system \cite{xiong2022characteristic} as well as some qualitative analysis results \cite{li2017wigner, ilivsevic2015stability}.

The rest of the paper is organized as follows. Section~\ref{sec:dirac} presents the numerical results for the Dirac delta function potential and an comparison with the finite size model. Scattering of the Fermi-Dirac distribution in 4-D phase space is shown as well. Extensions to other three types of singular potentials are given in Section~\ref{sec:exte}.
Finally,  conclusions and discussions are drawn in Section~\ref{sec:conc}.

\section{Quantum dynamics in a Dirac delta function potential}
\label{sec:dirac}

After truncating the $k$-space into $\mathcal{K} = [ k_{min}, k_{max}]$ \cite{xiong2016advective}, 
a Fourier spectral approximation with $N_k$ terms to the Wigner function $f(x, k, t)$ reads 
\begin{equation}\label{fk_plane}
f(x,k,t) \approx f_{N_k}(  x,  k, t)=\sum_{\nu=-N_k/2 + 1}^{N_k/2}
\alpha_{\nu}(  x, t)  \, \psi_{\nu}(k),  \
\end{equation}
where $\psi_{\nu}(k)=e^{2 \pi \mi \nu (k - k_{min})/L_k}$ with $L_k = k_{max} - k_{min}$ gives the basis.
Then the pseudo-differential term \eqref{Th} can be approximated as follows
\begin{align}
 \Theta_{V}[f](  x,   k, t)  &\approx  \Theta^T_{V}[f_{N_k}](  x,   k, t)  = \sum_{\nu=-N_k/2 + 1}^{N_k/2}  c_\nu( x)\, \alpha_{\nu}(  x, t)\, \psi_{\nu}(k), \label{thetaT} \\
c_\nu( x) &= \int_{\mathcal{K^\prime}} V_w (  x,   k^\prime)\,  \me^{-2\pi \mi \nu k^\prime/L_k} \, \dif  k^\prime,
\quad \mathcal{K^\prime} = [-L_k, \, L_k]. \label{c}
\end{align}
For the situation of the Dirac delta function potential~\eqref{delta-pot}, plugging Eq.~\eqref{Vw_delta} into Eq.~\eqref{c} leads to the following close formula 
\begin{equation}
\label{c_delta}
c_\nu( x) = \frac{2H\mi}{\pi\hbar}\left(\frac{\sin(\omega^+_\nu(x) L_k)}{\omega^+_\nu(x)}
           - \frac{\sin(\omega^-_\nu(x) L_k)}{\omega^-_\nu(x)}\right),
\end{equation}
where $\omega^\pm_\nu(x) = 2x\pm\frac{2\pi}{L_k}\nu$ and the limits must be used when $\omega^\pm_\nu(x)=0$. 
It can be easily observed in Eq.~\eqref{thetaT} that only $c_\nu( x)$ involves the singular potential \eqref{delta-pot}, through the non-singular Wigner kernel \eqref{Vw_delta}, thereby implying that the formula \eqref{c_delta} treats the singularity with high accuracy, where the numerical errors only come from the truncation of $k$-space and the spectral approximation of  $f(x,k,t)$. After that, we adopt the Chebyshev spectral element method with inflow boundary conditions \cite{shao2011adaptive} 
in $x$-space and the fourth-order operator splitting technique \cite{2019A} in $t$-direction to
determine the remaining unknowns $\alpha_{\nu}(  x, t)$. 
For simplicity, we use the same $M$ collocation points in all $Q$ cells in $x$-space. 
Moreover, the above numerical method can be readily extended to 4-D and higher-dimensional scenarios in a dimension-by-dimension manner by using the tensor product of 2-D basis functions. 

The $L^2$-error
\begin{equation}
\label{l2}
\epsilon_2(t) = \left[    \iint_{\mathcal{X}\times \mathcal{K}}  \left(F(x,k,t)  - f^{\text{ref}}(x,k,t) \right)^2 \dif x \dif k   \right]^{1/2} 
\end{equation}
and $L^\infty$-error
\begin{equation}
\label{lf}
\epsilon_\infty(t) =  \max_{(x,k)\in \mathcal{X}\times \mathcal{K}}  \left\{ |F(x,k,t)  - f^{\text{ref}}(x,k,t)| \right\}
\end{equation}
are used to analyze the convergence of the errors, where $\mathcal{X}:= [X_L, X_R]$ is the computational domain in $x$-space, $F(x,k,t)$ represents the numerical solution, and the numerical solution obtained on the finest mesh is taken as the reference $f^{\text{ref}}(x,k,t)$. 
For convenience, the above errors in Eqs.~\eqref{l2} and \eqref{lf} are numerically calculated on the following uniform mesh 
\begin{equation}
\label{Mesh_um}
(x_i, k_j) = ((i-{1}/{2}) {(X_{R} - X_{L})}, (j-{1}/{2})(k_{max} - k_{min}))/{N_{um}}, \ i, j=1,\dots,N_{um},
\end{equation}
where $N_{um}$ denotes the mesh size.

\subsection{2-D scattering of Gaussian wave packet}

As  stated in \cite{shao2011adaptive, xiong2016advective}, the Gaussian wave packet
\begin{equation}
\label{GaussF}
f(x, k, 0) =  \frac{1}{\pi}  \, e^{  -\frac{(x - x^0)^2}{2 \sigma^2}  - 2\sigma^2  (k - k^0)^2 }
\end{equation}
is usually adopted as the initial function to test the convergence rate as well as to investigate the quantum tunneling,  where $(x^0, k^0)$ gives the initial center position and $\sigma$ is  the minimum position spread. We will simulate its quantum scattering in the Dirac delta function potential, which has never been reported before in the literature. For the purpose of testing only, we set $\hbar = 1  ~\text{eV} \cdot \text{fs}$, $m = 1  ~\text{eV} \cdot \text{fs}^2 \cdot \text{nm}^{-2}$,  $x^0 =-10  ~\text{nm}$, $k^0=2  ~\text{nm}^{-1}$, and $\sigma=2  ~\text{nm}$. The computational domain is chosen as $[X_L, X_R]=[-30~\text{nm}, 30~\text{nm}]$ which is divided evenly into $Q=20$~cells, $-k_{min}=k_{max}=\pi  ~\text{nm}^{-1}$, and the quantum evolution with a fixed time step $\Delta t = 0.01$~fs is stopped at $t_{fin}=10$~fs.


\begin{figure}[htbp!]
\centering
\subfigure[$t=2.5 \,\text{fs}$.]
{\includegraphics[width=0.495\textwidth,height=0.38\textwidth]{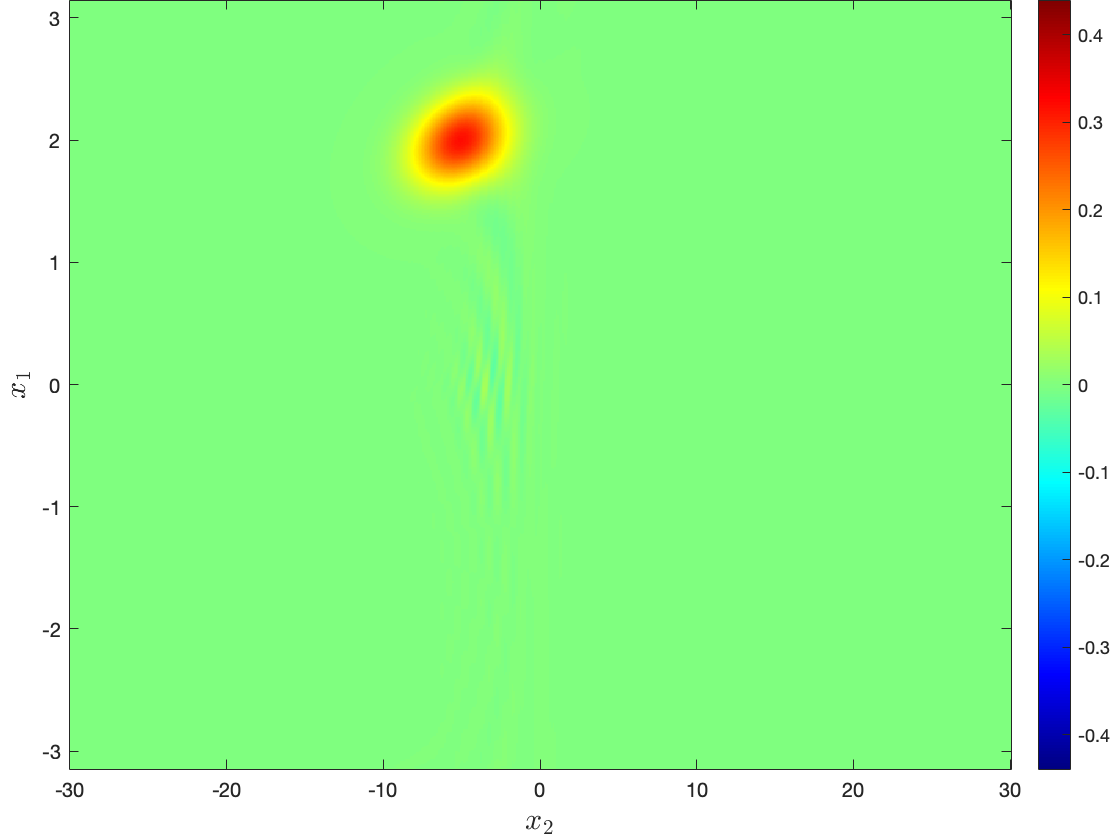}}
\subfigure[$t=5 \,\text{fs}$.]
{\includegraphics[width=0.495\textwidth,height=0.38\textwidth]{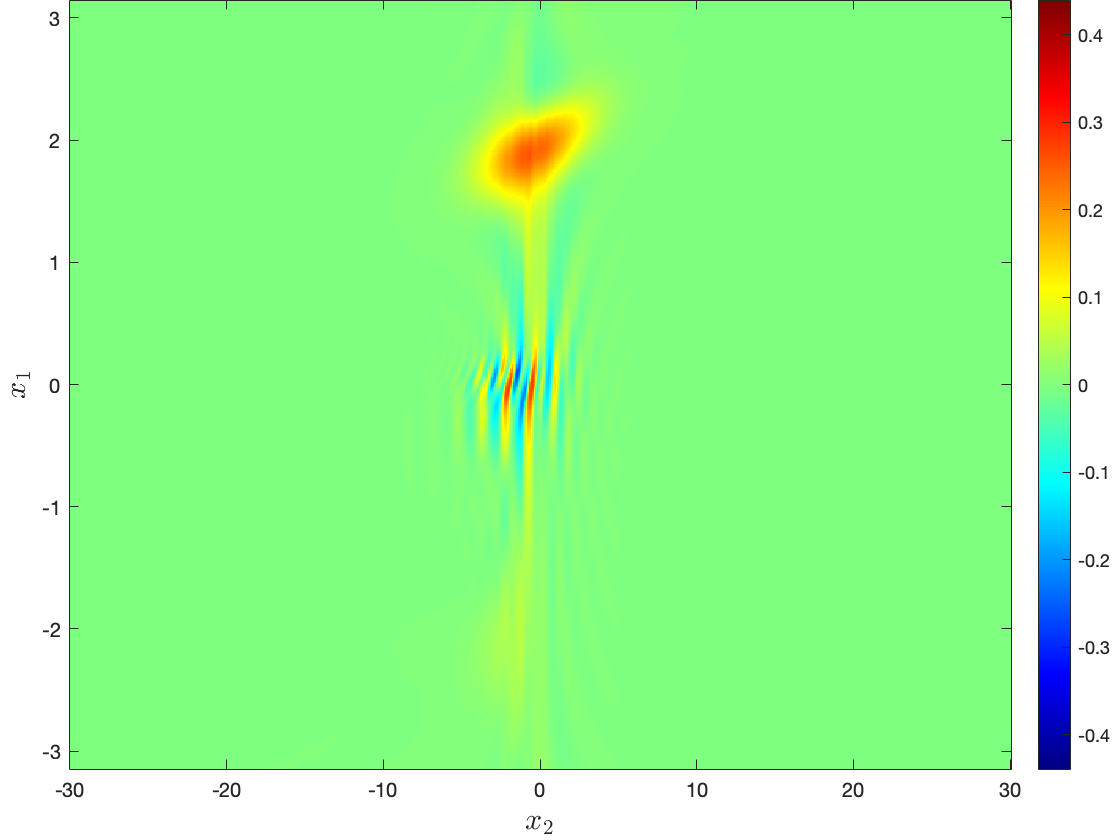}}\\
\subfigure[$t=7.5 \,\text{fs}$.]
{\includegraphics[width=0.495\textwidth,height=0.38\textwidth]{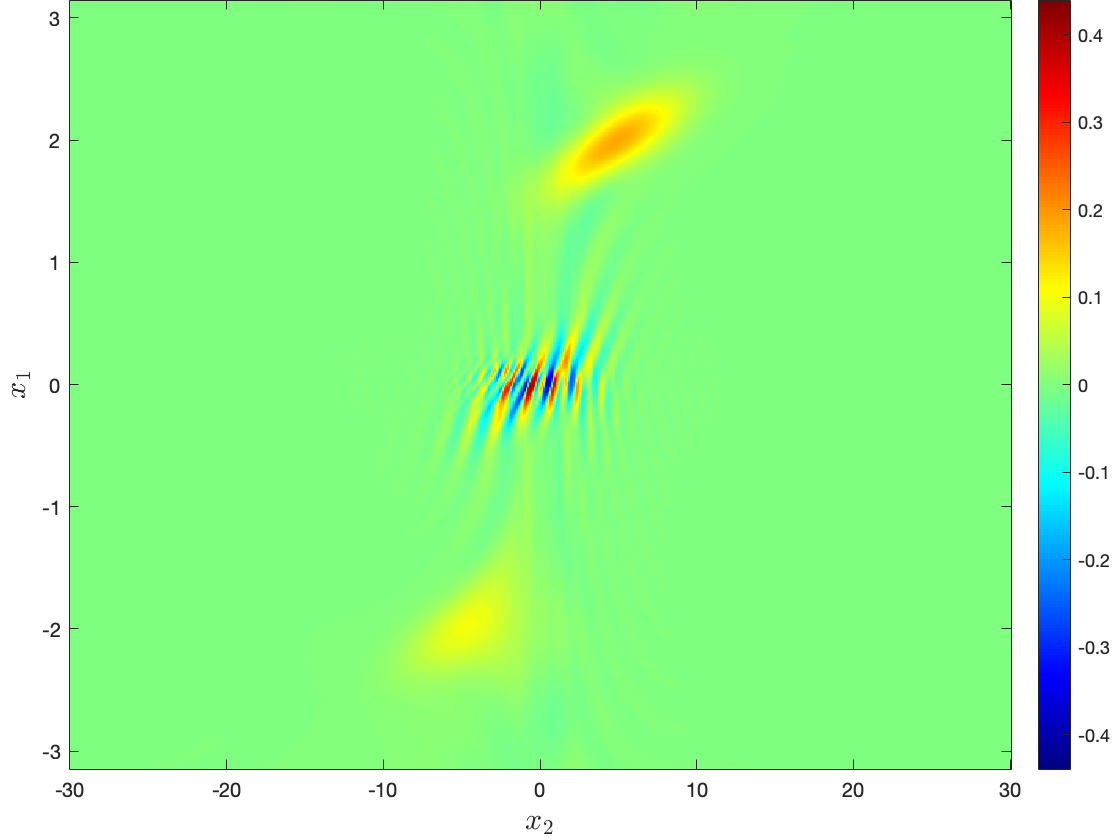}}
\subfigure[$t=10 \,\text{fs}$.]
{\includegraphics[width=0.495\textwidth,height=0.38\textwidth]{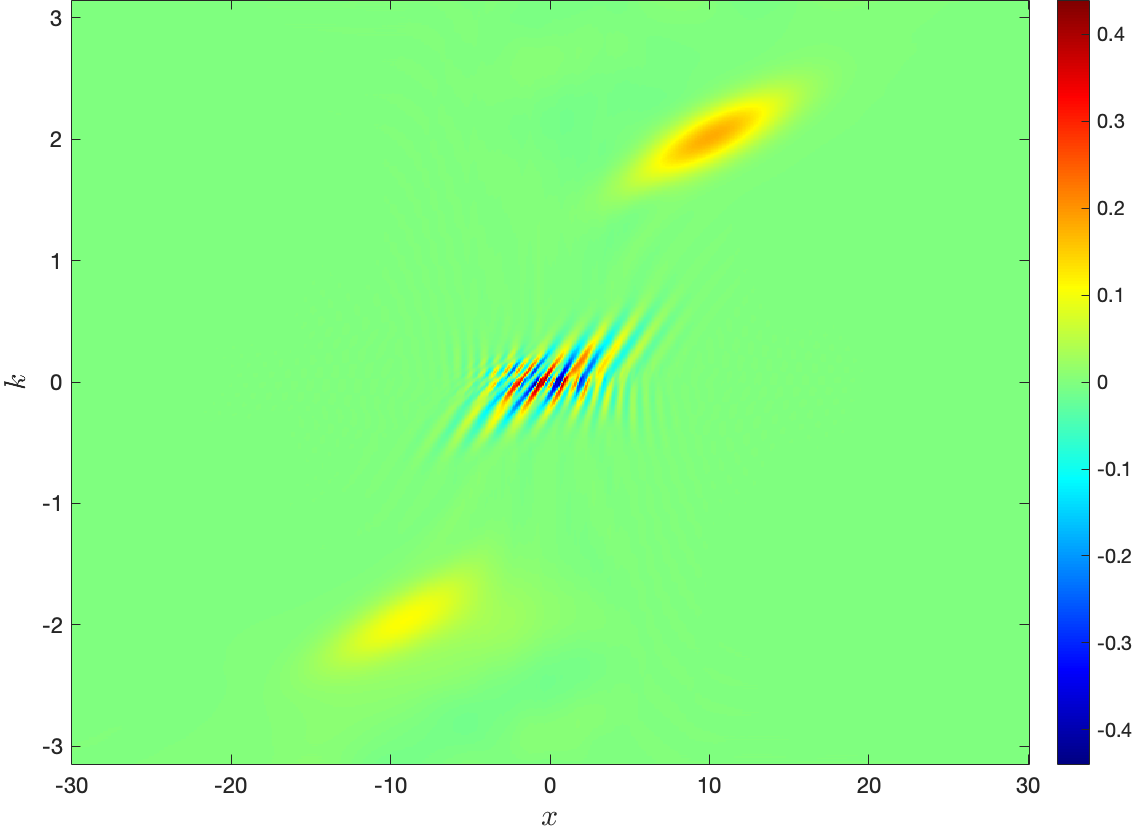}}
\caption[delta1]{\small The Wigner functions at different instants: A Gaussian wave packet runs through the Dirac delta function barrier Eq.~\eqref{delta-pot} with power $H=1$. Quantum tunneling effect can be clearly observed.}
\label{fig:WF_delta}
\end{figure}

\begin{figure}[htbp!]
	\centering 
\subfigure
{\includegraphics[width=0.495\textwidth,height=0.38\textwidth]{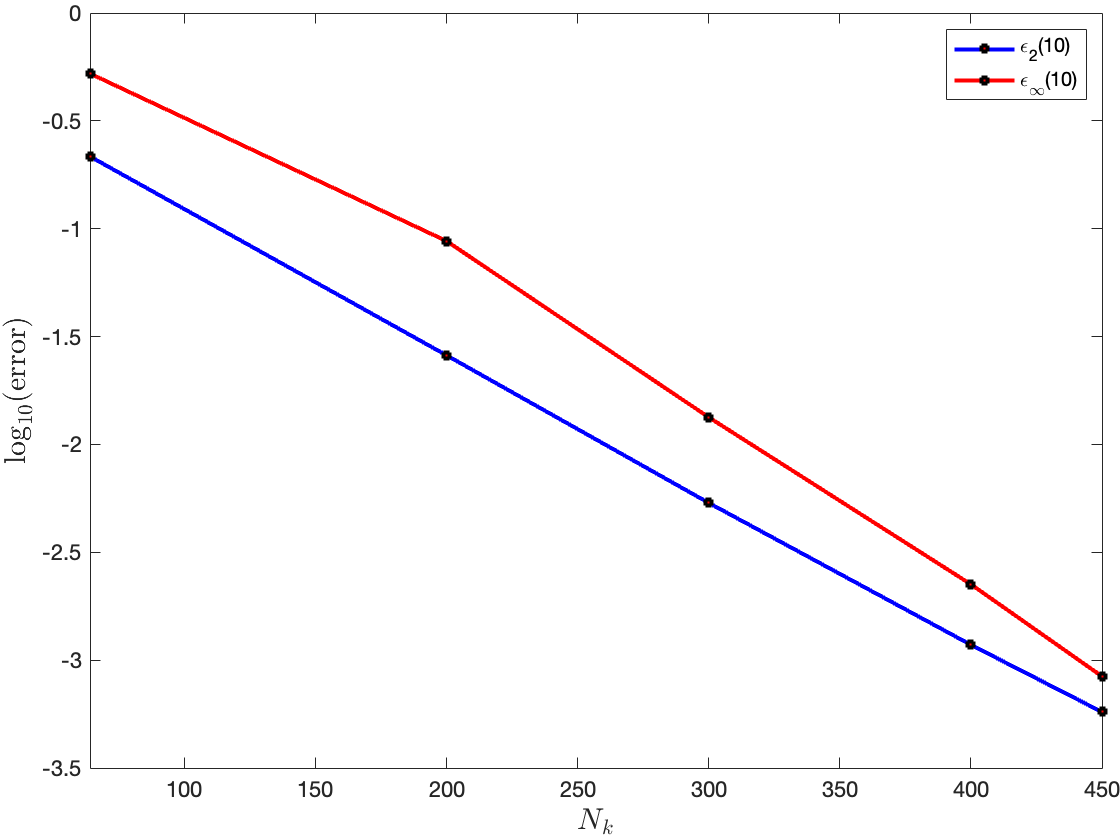}}
\subfigure
{\includegraphics[width=0.495\textwidth,height=0.38\textwidth]{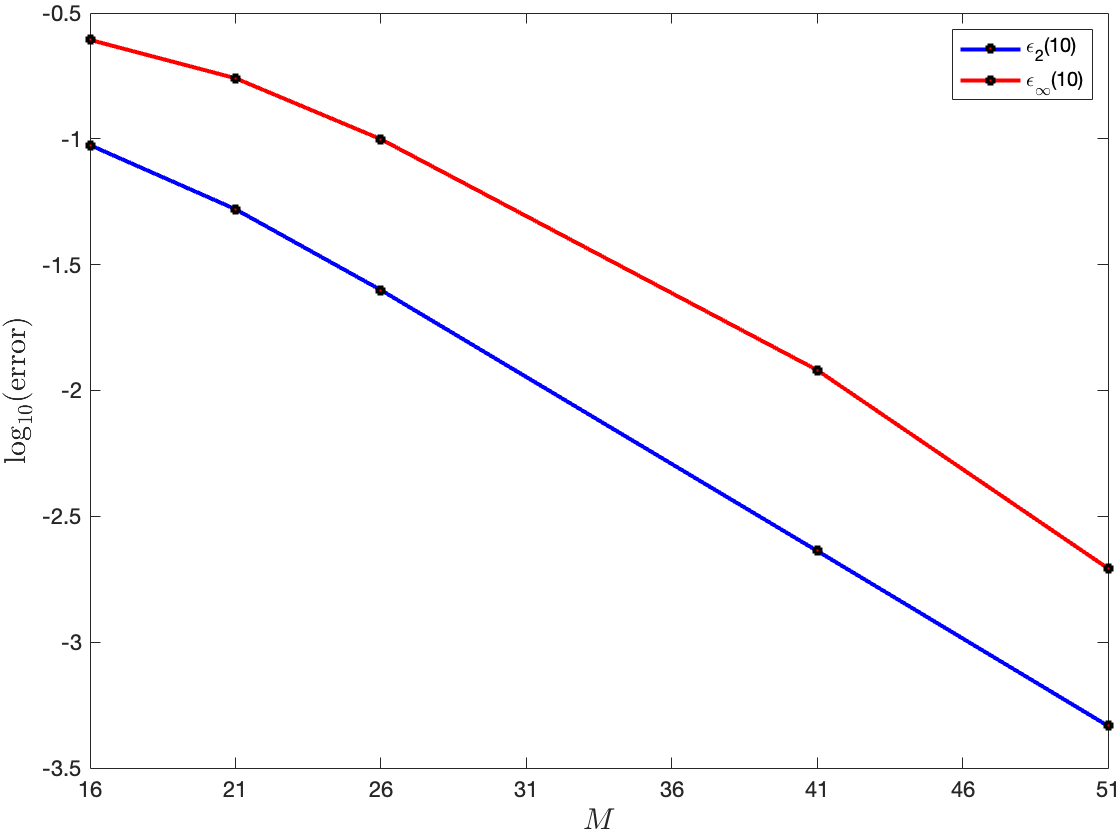}}
\caption[Convergence order delta]{\small Spectral convergence with respect to $N_k$ (left) and $M$ (right)
during the scattering of a Gaussian wave packet in the Dirac delta function potential Eq.~\eqref{delta-pot} with $H=1$. } 
\label{fig:delta_error} 
\end{figure}

The first test tries $H = 1$ in Eq.~\eqref{delta-pot} and the resulting Wigner functions at $t=2.5, 5, 7.5, 10$~fs are displayed in Fig.~\ref{fig:WF_delta}, which is obtained on the finest mesh we have tried: $N_k=512, M=55$.  It is clearly observed there that the wave packet still goes partially through the barrier though the barrier whose height is infinite. That is a clear manifestation of quantum tunneling effect in the Dirac delta function potential \cite{grabert1985quantum}, reflecting a fundamental difference between quantum world and macroscopic world. A possible explanation is that the width of the Dirac delta function barrier tends to be extremely small, very close to $0$ in particular, albeit the infinitely large height.
To study the convergence rate with respect to $N_k$, the number of collocation points in each $x$-cell is
fixed to be $M = 55$. Similarly, when studying the convergence rate with respect to $M$, the number of collocation points in $k$-space is fixed to be $N_k = 512$.  Fig.~\ref{fig:delta_error} displays the spectral convergence of $\epsilon_2(10)$ and $\epsilon_\infty(10)$ against $N_k$ or $M$ where we have set $N_{um}=600$ in Eq.~\eqref{Mesh_um}. That is, the numerical results in Fig.~\ref{fig:WF_delta} are numerically converged.

To be more specific, the uncertainty
\begin{equation}
 \sigma_x(t) \, \sigma_p (t) = \sqrt{\left< \left( \hat{x} - \left<\hat{x}\right>_t \right)^2 \right>_t}\, \sqrt{\left< \left( \hat{p} - \left<\hat{p}\right>_t \right)^2 \right>_t}
 \label{uncertainty}
 \end{equation}
is adopted to measure the nonlocality, and its numerical value is still obtained on the uniform mesh given in Eq.~\eqref{Mesh_um}, where $p=\hbar k$ is the momentum. 
We show in Table~\ref{tab:convergence_delta_uncertainty} that the numerical results with different mesh sizes and find that  the numerically converged value for $\sigma_x(10) \sigma_p(10)$ is $20.8510$ for $H=1$.

\begin{table}[!h]
  \centering
  \caption{\small   Numerical values for $\sigma_x(10) \sigma_p(10)$  with respect to increasing $N_k$, $M$ and $N_{um}$ and the converged value is $20.8510$.}
\label{tab:convergence_delta_uncertainty}
 \begin{lrbox}{\tablebox}
 \begin{tabular}{cc|cc|cc}
 \hline\hline
 $N_k$ & $\sigma_x(10) \sigma_p(10)$   & $M$ & $\sigma_x(10) \sigma_p(10)$  &   $N_{um}$ & $\sigma_x(10) \sigma_p(10)$ \\\hline
$32$   & $19.9616$ & $21$  & $21.0169$  & $100$  & $20.8355$   \\
$64$   & $20.8938$ & $26$  & $20.8682$  & $200$  & $20.8467$   \\
$128$ & $20.8570$ & $31$  & $20.8406$  & $300$  & $20.8510$  \\
$256$ & $20.8529$ & $36$  & $20.8502$  & $400$  & $20.8510$   \\
$300$ & $20.8524$ & $41$  & $20.8515$  & $450$  & $20.8510$  \\
$400$ & $20.8515$ & $45$  & $20.8509$  & $500$  & $20.8510$  \\
$500$ & $20.8510$ & $51$  & $20.8510$  & $550$  & $20.8510$  \\          
$512$ & $20.8510$ & $55$  & $20.8510$  & $600$  & $20.8510$  \\      
\hline\hline
 \end{tabular}
\end{lrbox}
\scalebox{1.0}{\usebox{\tablebox}}
\end{table}

\begin{figure}[htbp!]
\centering 
\subfigure[Tunneling rate.]
{\includegraphics[width=0.495\textwidth,height=0.38\textwidth]{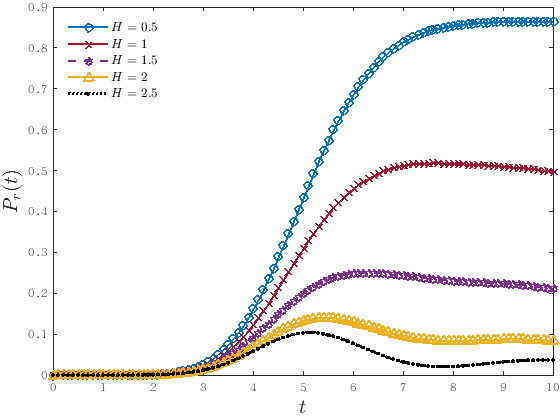}}
\subfigure[Uncertainty.]
{\includegraphics[width=0.495\textwidth,height=0.38\textwidth]{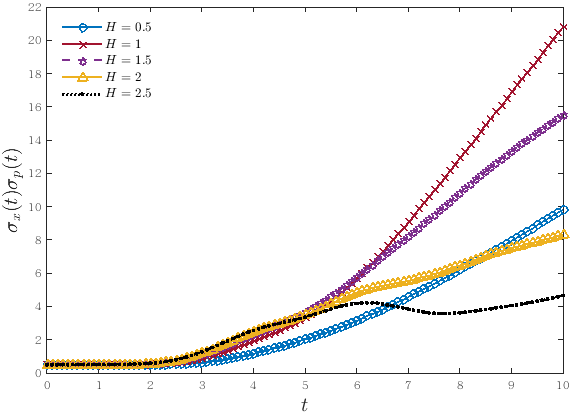}}
\caption[Composi\_Physical.]{ \small Tunneling rates and uncertainties for the Dirac delta function barriers with powers $H=0.5$, $1$,  $1.5$, $2$ and $2.5$. The moment when the uncertainty reaches the maximum coincides with that the tunneling rate reaches about $50\%$. } 
\label{fig:phy_deposi} 
\end{figure}

\begin{figure}[ht!]
\centering 
\subfigure[Tunneling rate.]
{\includegraphics[width=0.495\textwidth,height=0.38\textwidth]{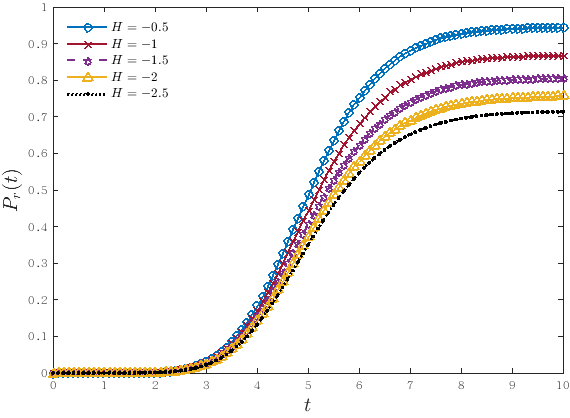}}
\subfigure[Uncertainty.]
{\includegraphics[width=0.495\textwidth,height=0.38\textwidth]{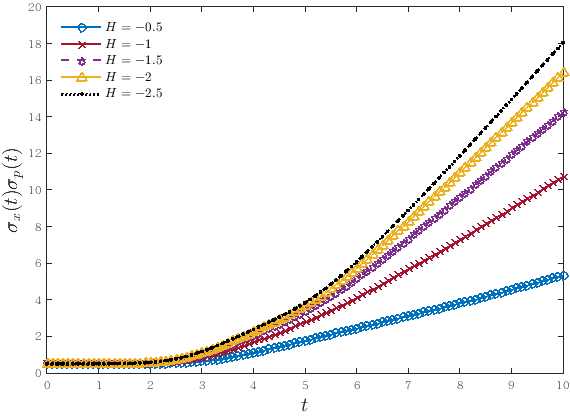}}
\caption[Comnega\_Physical.]{\small Tunneling rates  and uncertainties for the Dirac delta function well with powers $H=-0.5$, $-1$,  $-1.5$, $-2$ and $-2.5$.} 
\label{fig:phy_denega} 
\end{figure}

In addition to the uncertainty, we continue to use the partial mass
 \begin{equation}
 \label{Pr}
 P_r (t) = \int_{[0,X_R] \times \mathcal{K}} f(x,k,t)~\dif x \dif k
\end{equation}
for investigating the tunneling effect for ten different powers: $H = \pm 0.5$, $\pm 1$, $\pm 1.5$, $\pm 2$, and $\pm 2.5$. $P_r$ can also be regarded as the tunneling rate in view of that the total mass equals to one.
 Figs.~\ref{fig:phy_deposi} and \ref{fig:phy_denega} show the tunneling rates and uncertainties for potential barriers $H>0$ and potential wells with $H<0$, respectively. 
The curves in Fig.~\ref{fig:phy_deposi} exhibit the deceleration of $P_r(t)$ as $H$ increases, and uncertainty peaking at $H=1$. 
It can be  further found that the moment when the uncertainty reaches the maximum coincides with that the tunneling rate reaches to about $0.5$. At this moment, the variances accumulate the most  and it is difficult to observed the position and momentum of the wavepacket simultaneously. 
Moreover, when the power is high enough, i.e., $H \ge 1.5$,   there are significant fluctuations of $\sigma_x(t) \, \sigma_p(t)$ and $P_r (t)$. That could be comprehended as follows. The influence of the power $H$ leads to the high oscillations in the Wigner function at the center $x=0$~nm. Therefore, these two observables fluctuate during the wave packet interacting with the barrier.
When it comes to the wells with negative power, it can be observed in Fig.~\ref{fig:phy_denega}  that the trend of the tunneling rate $P_r(t)$ is opposite to that of the uncertainty $\sigma_x(t) \, \sigma_p (t)$: 
$P_r(t)$ decreases and $\sigma_x(t) \, \sigma_p (t)$ increases as as $|H|$ increases.

\subsection{Finite size effect}
The point charge causes singularity because it has no size. In view of this,
one may use a finite size model to avoid the singularity. 
The Gaussian function with size of  $a$, denoted by $V_a(x)$, 
is usually used to mimic the point charge model \cite{doi:10.1063/1.2736702, visscher1997dirac}, 
the validity of which relies on the following limit 
\begin{equation}
\label{V_gauss}
\delta(x) = \lim_{a\to 0+0} V_a(x) = \lim_{a\to 0+0} \frac{1}{ \sqrt{2\pi} a} \exp{(- \frac{x^2}{2 a^2} )}.
\end{equation}
However, we would like to point out that there is a huge gap between the quantum behavior caused by the point charge $\delta(x)$ and finite-size charge $V_a(x)$.

\begin{table}[!h]
  \centering
  \caption{\small Uncertainties for different sizes at $t_{fin}=10$~fs. 
  $a =0$~nm signifies the Dirac delta function barrier. The uncertainty of the Dirac delta function barrier $\delta(x)$ is much larger than that of the Gaussian barrier of any finite size $V_a(x)$.}
\label{tab:uncertainty}
 \begin{lrbox}{\tablebox}
 \begin{tabular}{cccccccccc}
 \hline\hline
 $a$~nm & $10$ &  $5$  & $0.5$ &  $0.1$ & $0.01$ & $1$E-3 & $1$E-4 & $1$E-16 & $0$   \\
 \hline
$\sigma_x(10) \, \sigma_p(10)$	&  $0.9811$  & $1.0663$ & $1.2272$ &$1.0041$ & $0.8027$ &$0.8005$ &$0.8003$ & $0.8003$ & $20.8510$\\
\hline\hline
 \end{tabular}
\end{lrbox}
\scalebox{0.85}{\usebox{\tablebox}}
\end{table}


\begin{figure}[htbp!]
	\centering 
\subfigure[$V_a(x)$ with $a =0.5$~nm.]
{\includegraphics[width=0.495\textwidth,height=0.38\textwidth]{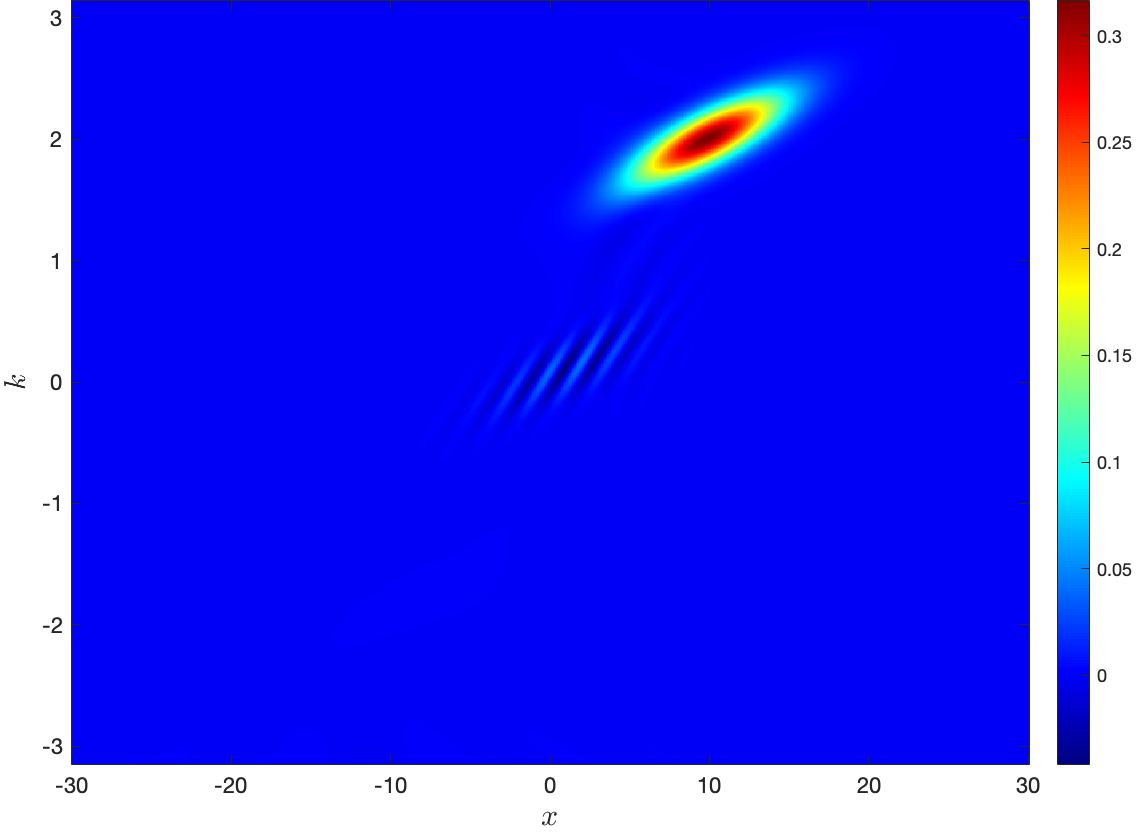}\label{fig:gaVSde:p1}}
\subfigure[$\delta(x)$.]
{\includegraphics[width=0.495\textwidth,height=0.38\textwidth]{./delta1_100.png}\label{fig:gaVSde:p2}}\\
\subfigure[Tunneling rate.]
{\includegraphics[width=0.495\textwidth,height=0.38\textwidth]{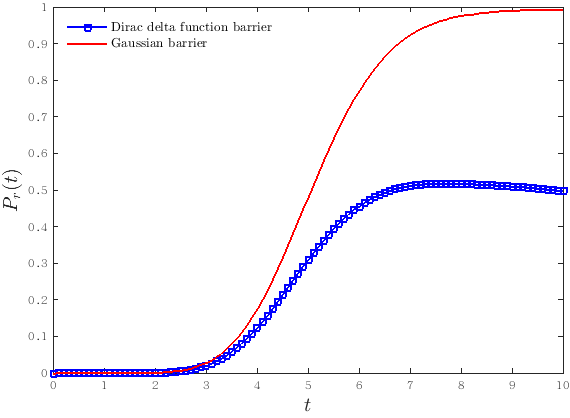}\label{fig:gaVSde:p3}}
\subfigure[Uncertainty.]
{\includegraphics[width=0.495\textwidth,height=0.38\textwidth]{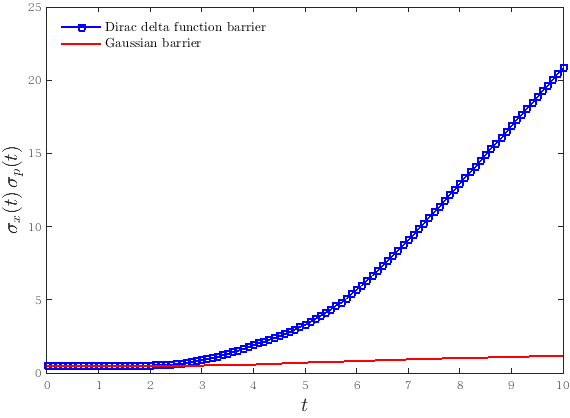}\label{fig:gaVSde:p4}}
\caption[deVSga.]{\small Different quantum behavior caused by the Gaussian barrier with finite size $V_a(x)$ and Dirac delta function barrier $\delta(x)$.  The Wigner functions at $t_{fin}=10$~fs are displayed in \subref{fig:gaVSde:p1} and \subref{fig:gaVSde:p2}, the tunneling rate in \subref{fig:gaVSde:p3}, and the uncertainty in \subref{fig:gaVSde:p4}. Both tunneling rates and uncertainties are obviously different. } 
\label{fig:gaVSde} 
\end{figure}

Table~\ref{tab:uncertainty} displays the numerically converged uncertainties at $t_{fin}=10 ~\text{fs}$ for decreasing sizes. When $a=10^{-16}$~nm, $\sigma_x(10) \, \sigma_p(10)$ is about $0.8003$, which is far less that $20.8510$ caused by $\delta(x)$. In fact, as the size gradually becomes smaller, the uncertainty first grows to $1.2272$, then gradually decreases, and finally stays around $0.8003$.
Such apparent discrepancy can be also observed from the Wigner functions at the final instant in Fig.~\ref{fig:gaVSde}. Compared the Wigner function under $V_a(x)$ with $a = 0.5$~nm in Fig.~\ref{fig:gaVSde:p1}, it is obvious that the Wigner function under $\delta(x)$ in Fig.~\ref{fig:gaVSde:p2} reaches the extrema around $x=0$~nm,  where the singular point exactly locates at,  and the oscillation between positive and negative values is more vicious. More specifically, at the final instant, the average position and moment are $( \left< \hat{x} \right>_{10}, \left< \hat{p}\right>_{10}  ) = (9.7732,1.9840)$ under the Gaussian barrier, but alter to $( \left< \hat{x} \right>_{10}, \left< \hat{p} \right>_{10} ) = (1.1313, 0.0047)$ under $\delta(x)$.
Fig.~\ref{fig:gaVSde:p3} provides the tunneling rate $P_r(t)$.  It reaches almost to $1$ as expected under the Gaussian barrier, which is consistent with the weak presence of negative part of the Wigner function in Fig.~\ref{fig:gaVSde:p1}. By contrast, it is manifested in Fig.~\ref{fig:gaVSde:p4} that the same wave packet can just partially pass through the Dirac delta function barrier and its uncertainty increases significantly,
which should result from the infinite height of the potential.

In a word, our numerical experiments suggests an essential difference between the singular potential and its regularized counterpart as already shown in investigating nuclear magnetic shielding \cite{doi:10.1063/1.2736702}. 
No matter how small the size of Gaussian barrier we choose, it is still a smooth and local potential.
The Dirac delta function potential, on the contrary, an ideal model widely used to simulate the point charge field source of quantum chemical reactions,  has some essentially difference from such regularized one in studying quantum phenomena.

\subsection{4-D scattering of Fermi-Dirac distribution}

\begin{figure}[htbp!]
\centering 
\subfigure
{\includegraphics[width=0.495\textwidth,height=0.38\textwidth]{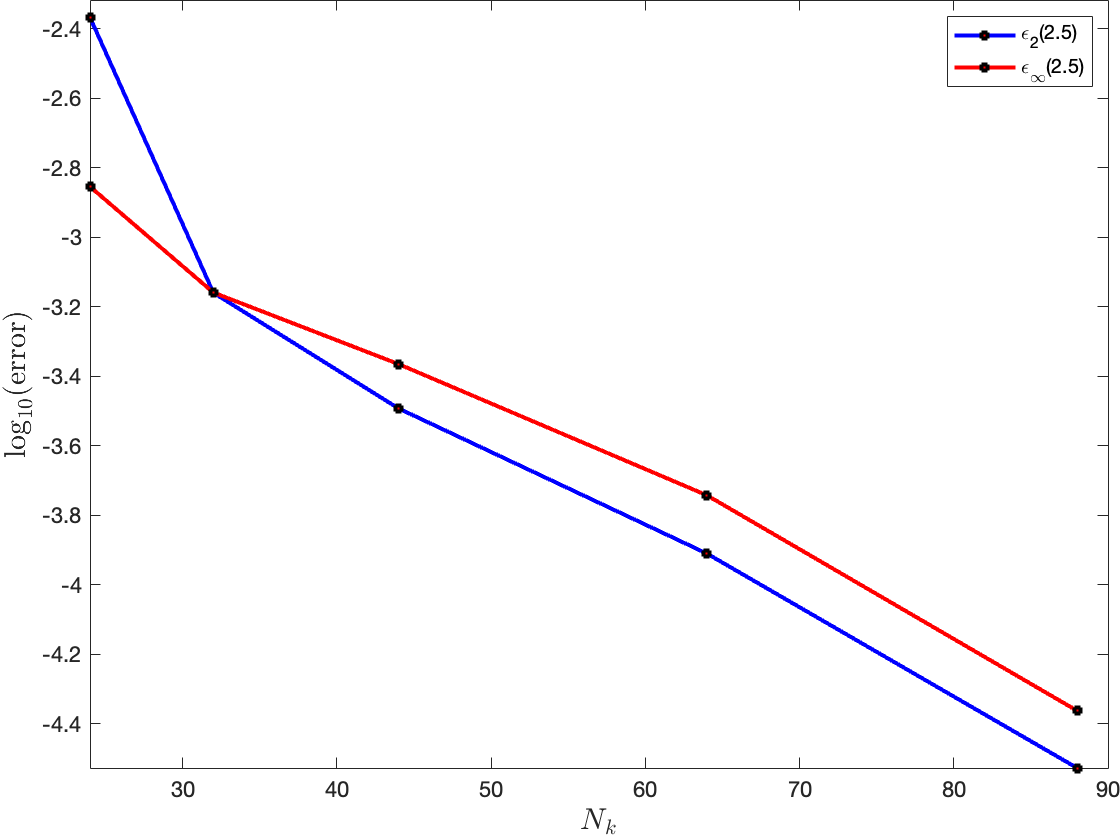}}
\subfigure
{\includegraphics[width=0.495\textwidth,height=0.38\textwidth]{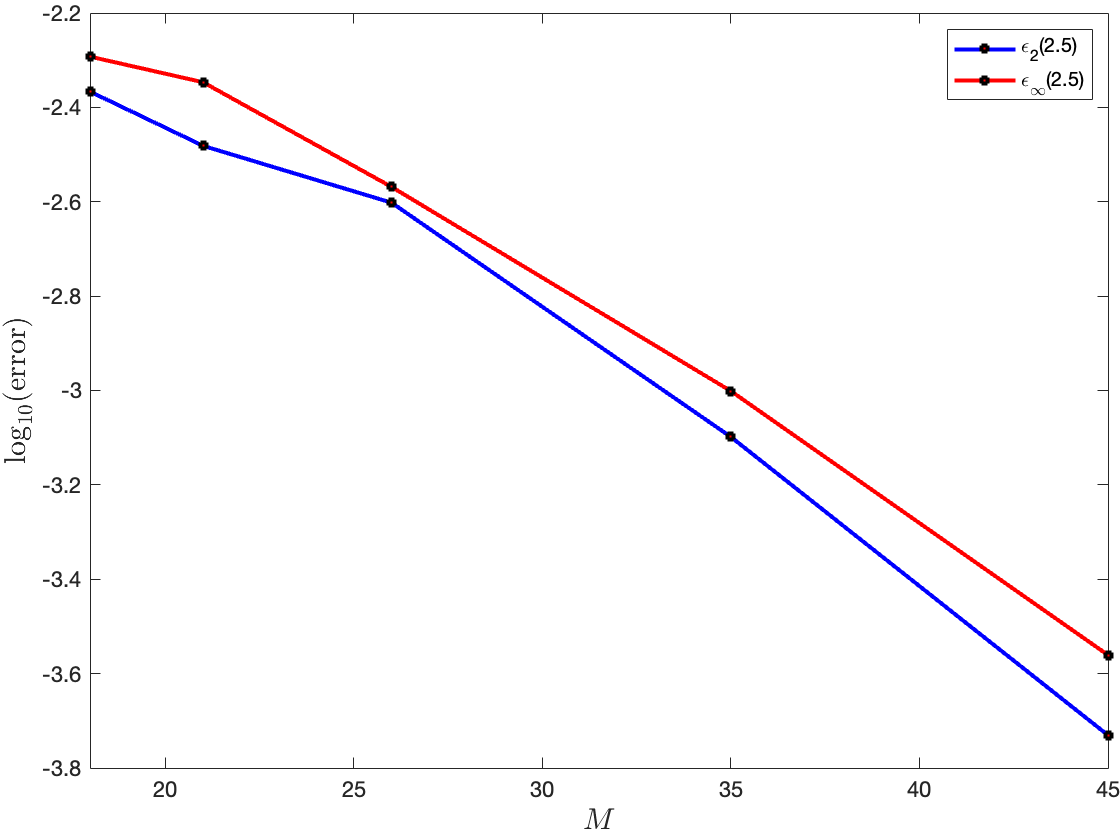}}
\caption{\small 4-D scattering of Fermi-Dirac distribution: Errors of the spatial marginal distribution
against $N_k$ (left) and $M$ (right).} 
\label{fig:2dDirac_error} 
\end{figure}

\begin{figure}[htbp!]
\centering
\subfigure[$t=0.1~\text{fs} $.]
{\includegraphics[width=0.328\textwidth,height=0.3\textwidth]{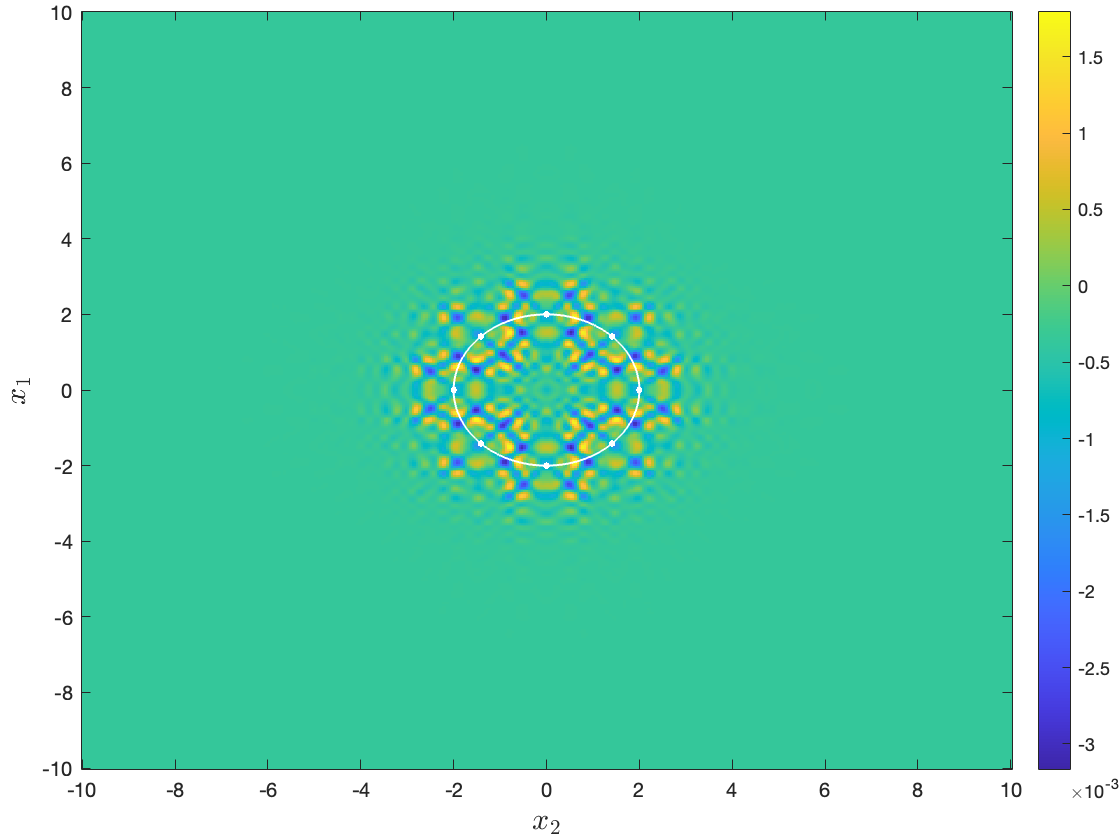}}
\subfigure[$t=0.2 ~\text{fs} $.]
{\includegraphics[width=0.328\textwidth,height=0.3\textwidth]{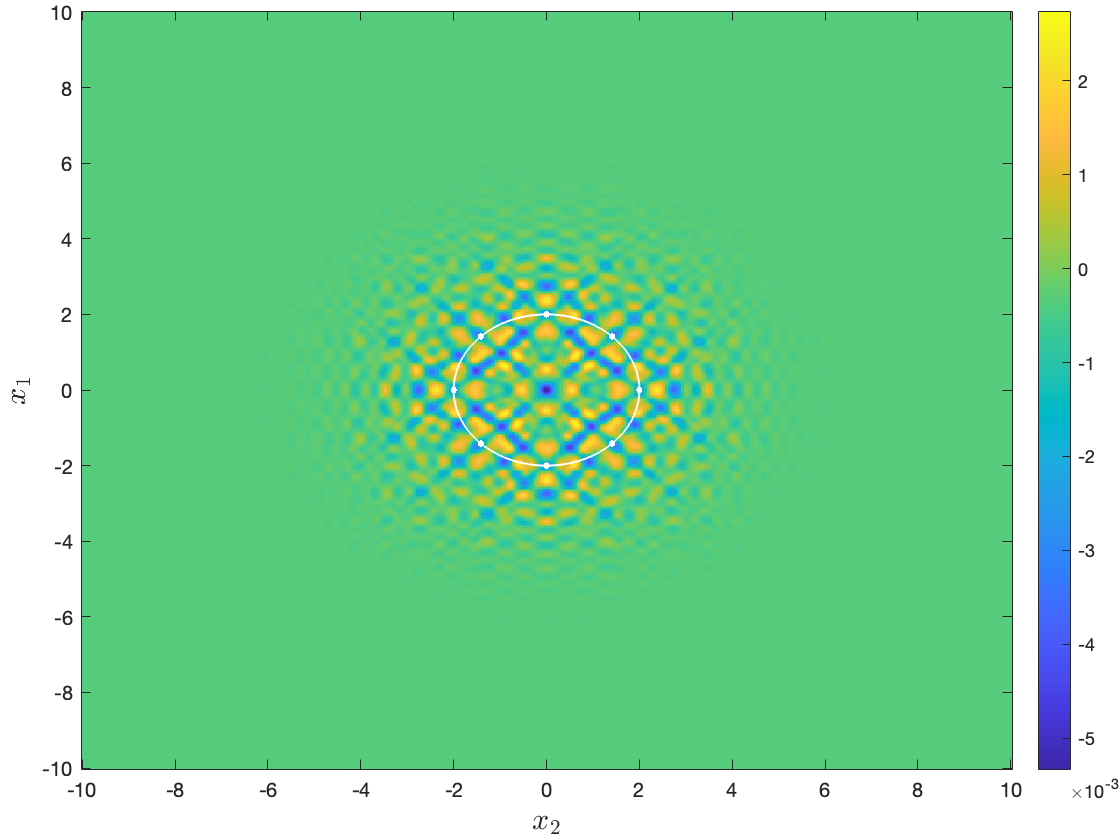}}
\subfigure[$t=0.5~\text{fs} $.]
{\includegraphics[width=0.328\textwidth,height=0.3\textwidth]{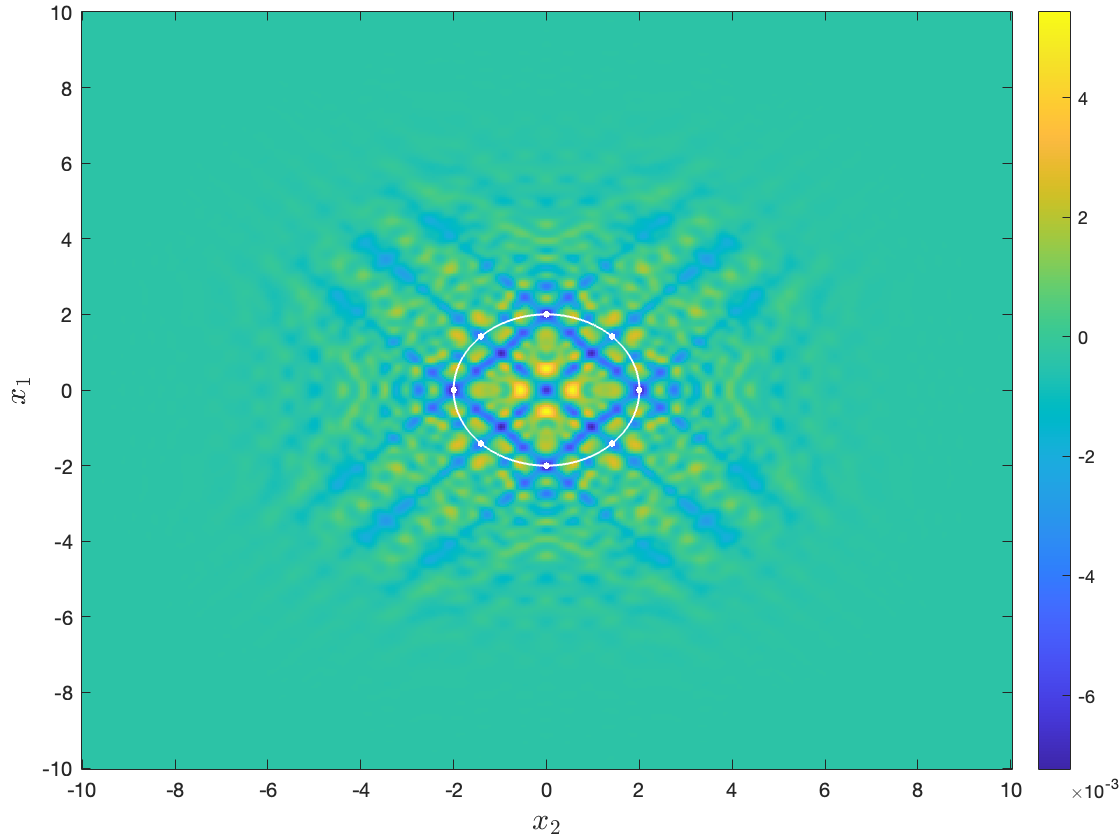}}\\
\subfigure[$t=0.8 ~\text{fs} $.]
{\includegraphics[width=0.328\textwidth,height=0.3\textwidth]{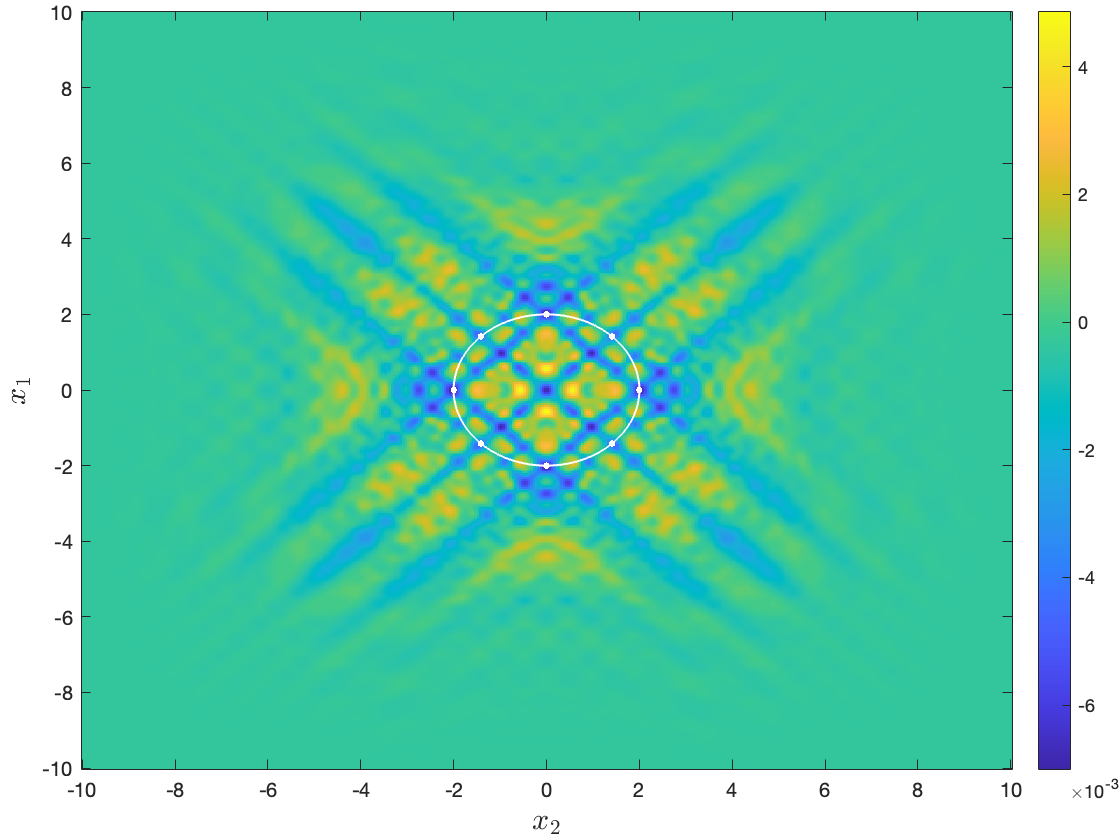}}
\subfigure[$t=1 ~\text{fs} $.]
{\includegraphics[width=0.328\textwidth,height=0.3\textwidth]{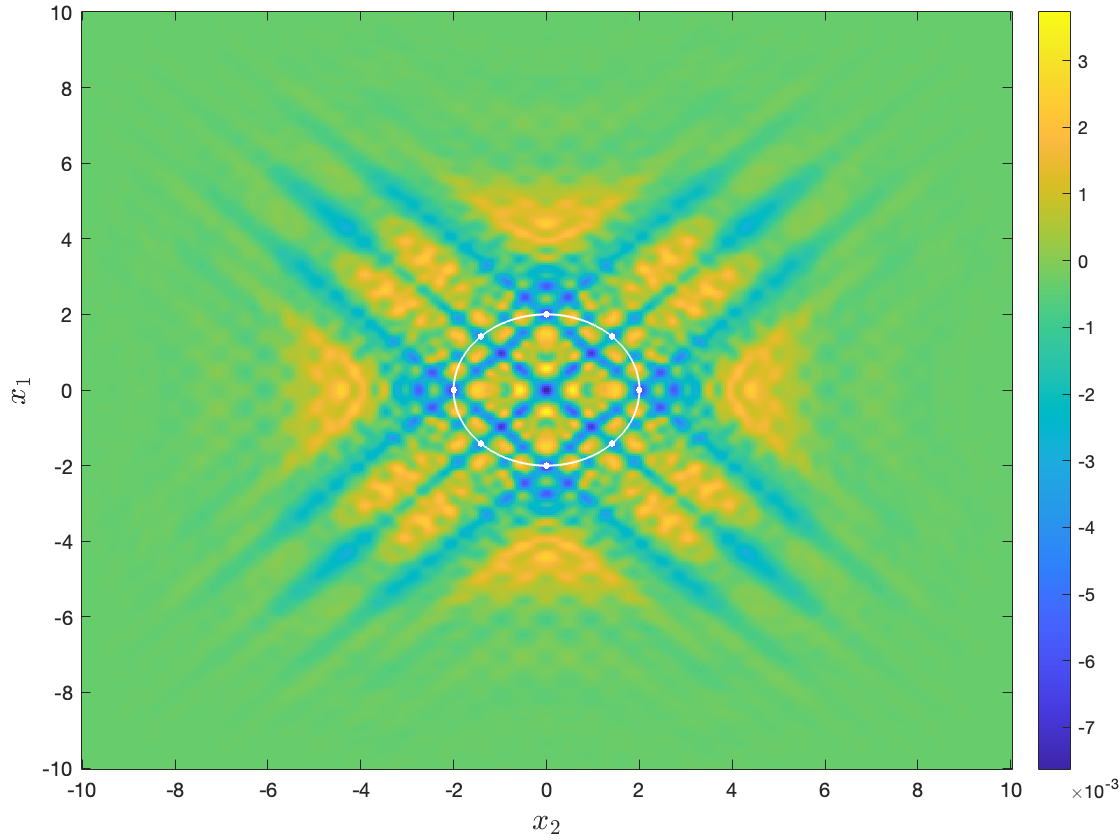}}
\subfigure[$t=1.5 ~\text{fs} $.]
{\includegraphics[width=0.328\textwidth,height=0.3\textwidth]{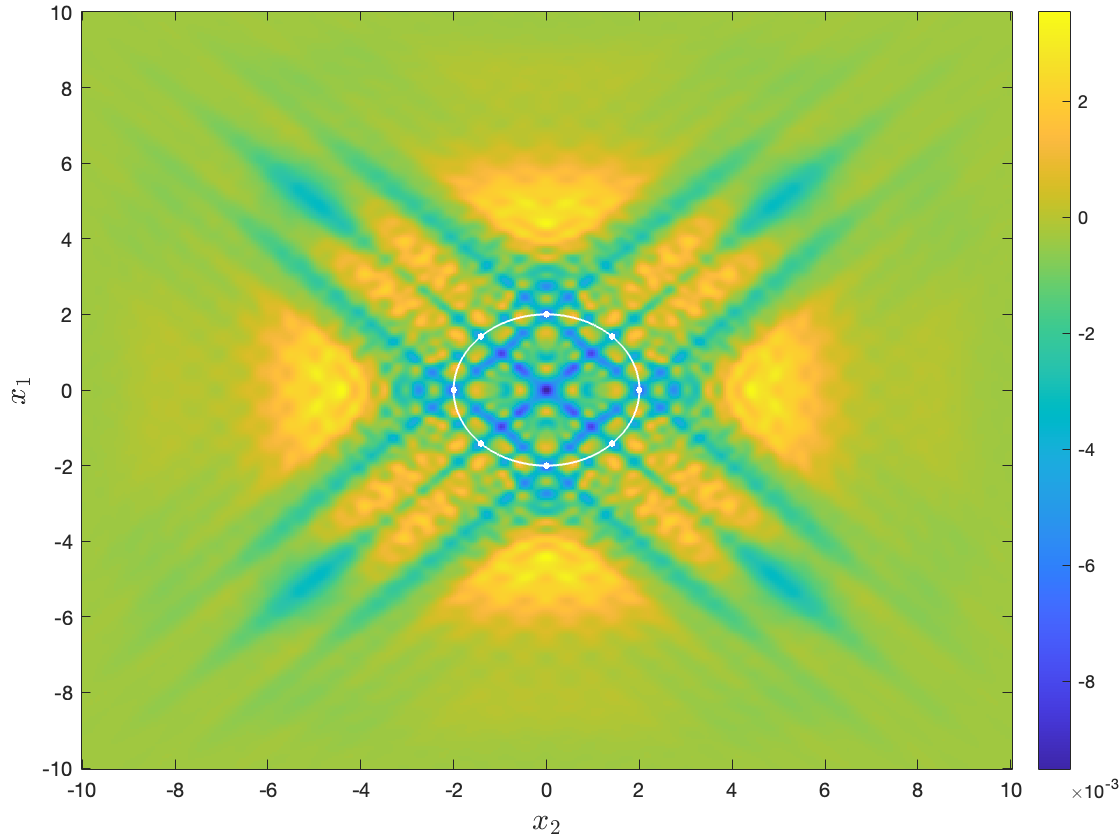}}\\
\subfigure[$t=2~\text{fs} $.]
{\includegraphics[width=0.328\textwidth,height=0.3\textwidth]{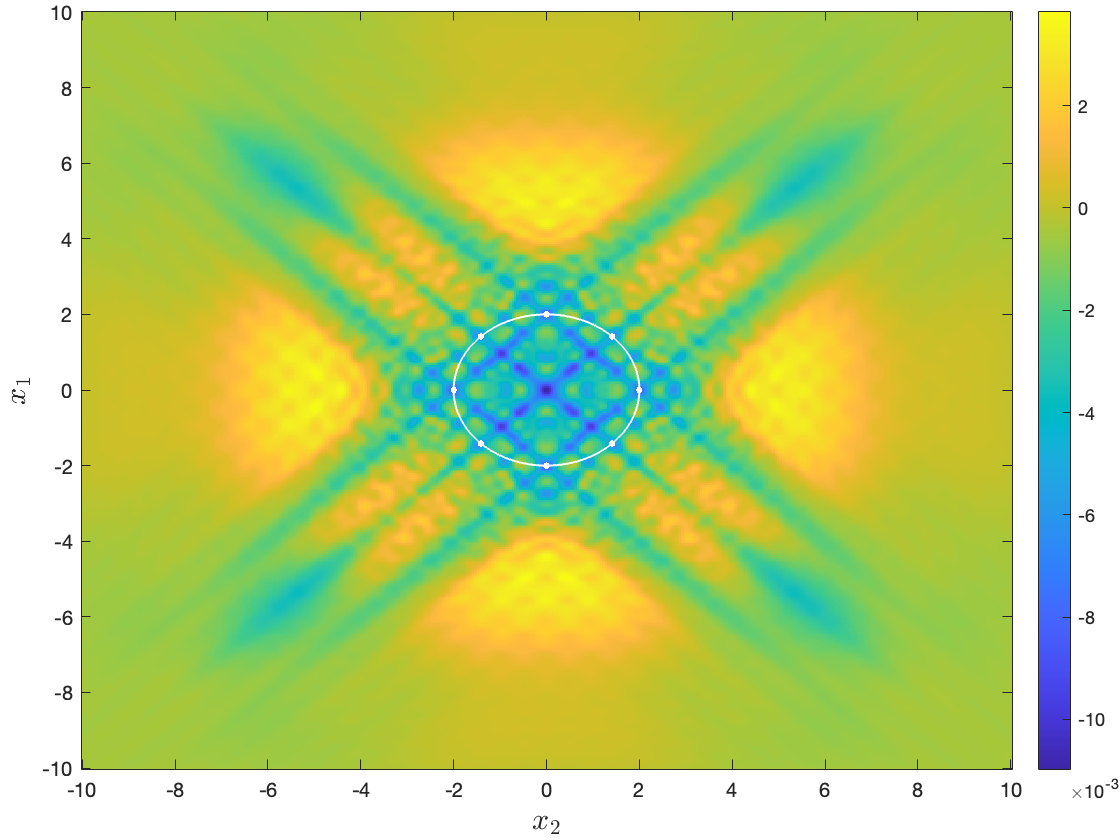}}
\subfigure[$t=2.5 ~\text{fs} $.]
{\includegraphics[width=0.328\textwidth,height=0.3\textwidth]{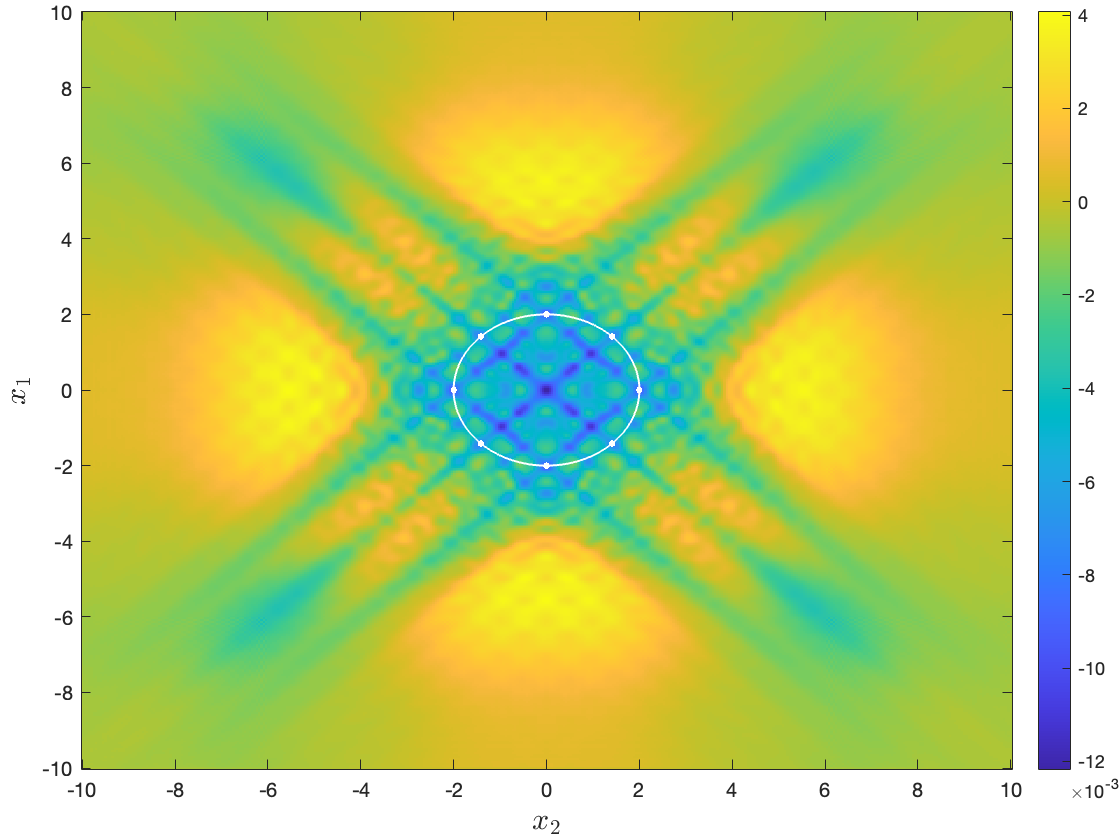}\label{fig:delta_fermidirac8fd2d_t25}}
\subfigure[Contour inside the circle.]
{\includegraphics[width=0.328\textwidth,height=0.3\textwidth]{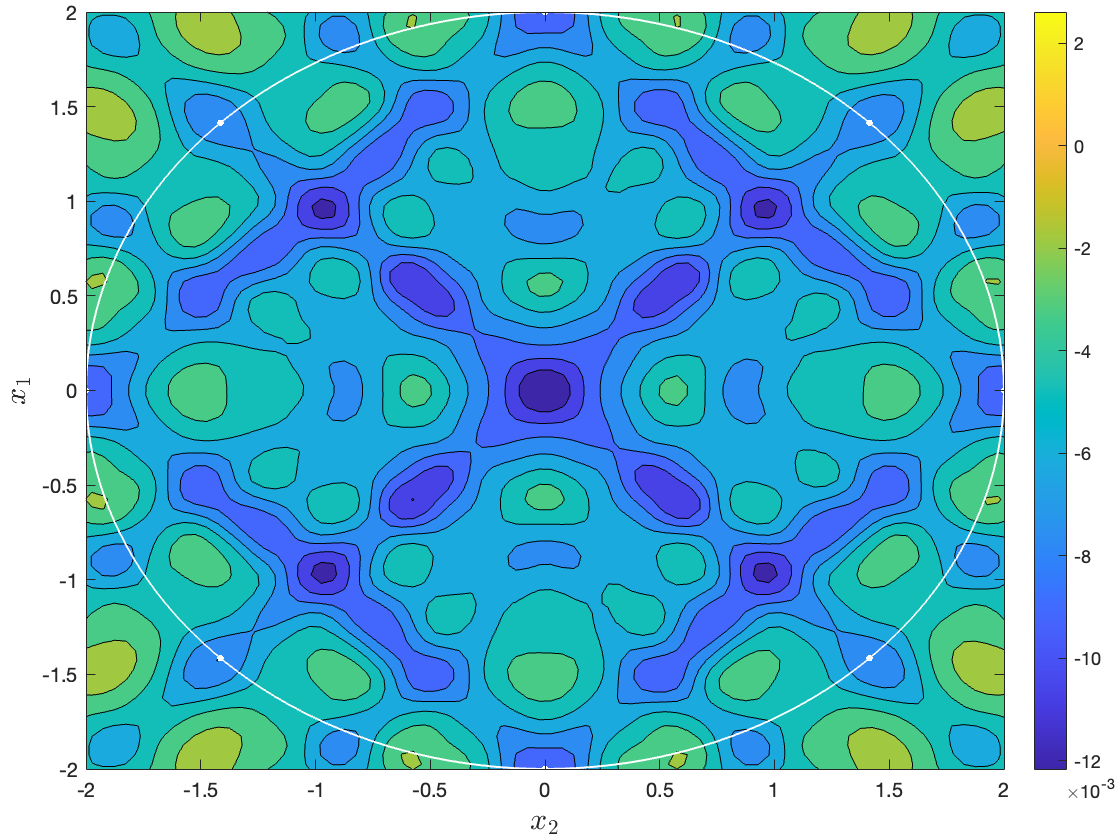}\label{fig:delta_fermidirac8fd2d_t25_contour}}\\
\caption{\small 4-D scattering of Fermi-Dirac distribution: Spatial marginal distributions in Eq.~\eqref{eq:fsm} subtracted by the corresponding constant distributions in the free space.  Eight Dirac delta function potentials give eight singular points (small white dots),  which are numbered $1$ to $8$ in an anti-clockwise direction with the right-most one being the first and evenly distributed in a circle with radius equal to $2$~nm. In \subref{fig:delta_fermidirac8fd2d_t25_contour}, we plot the contour inside the circle at $t=2.5 ~\text{fs}$ with $10$ equally spaced contour lines from $-0.01216$ to $0.004087$.}
\label{fig:delta_fermidirac82d}
\end{figure}

In view of the high dimensionality of phase space, the foregoing treatment of singular potentials using the Wigner function approach can be also extended to high-dimensional scenarios. This section is devoted to scattering of the Fermi-Dirac distribution in 4-D phase space under a singular potential. Specifically, we adopt the following 
position-independent 2-D Fermi-Dirac distribution function \cite{chen2022higher,macleod1998algorithm} as the initial data for the Wigner equation~\eqref{eq:WEQ}
\begin{equation}
\label{eq:fd0}
f(x_1, x_2, k_1, k_2, 0) = \frac{\sqrt{2m k_B T} }{\pi \hbar} \int_0^\infty \frac{1}{1+ \exp(y^2 + \frac{((\hbar k_1)^2 + (\hbar k_2)^2)/(2m) - E_{\text{F}}}{k_B T})} \dif y,
\end{equation}
where $m = 0.067\, m_e$, $m_e= 5.68562966 ~\text{eV}\cdot \text{fs}^{2} \cdot \text{nm}^{-2}$, 
$\hbar = 0.658211899  ~\text{eV} \cdot \text{fs}$,  $k_B = 8.61734279 \times {10}^{-5}  ~\text{eV} \cdot {\text{K}}^{-1}$,   $T$ is taken as $300  ~\text{K}$, and $E_{\text{F}}=0.1~\text{eV}$ signifies the Fermi energy. 
Meanwhile, we choose an annular singular potential $V(x_1, x_2) = \sum_{i=1}^8 \delta(x_1-d^i_1) \delta(x_2-d^i_2)$, where all singular points, numbered $1$ to $8$ in an anti-clockwise direction with the right-most one being the first, are evenly distributed in a circle with radius equal to $2$~nm and $(d_1^i, d_2^i)$ gives the position of the $i$-th singular point.

The computational domain is $\mathcal{X}\times\mathcal{X}\times\mathcal{K}\times\mathcal{K}$ with $\mathcal{X}=[-10~\text{nm}, 10~\text{nm}]$ and $\mathcal{K}=[-\pi  ~{\text{nm}}^{-1}, \pi ~{\text{nm}}^{-1}]$. We use the same $N_k$ collocation points for all $\mathcal{K}$,  the same number of elements $Q=5$ and $M$ collocation points in each element for all $\mathcal{X}$. 
The quantum dynamics is evolved to $t_{fin}=2.5~\text{fs}$ with time step $\Delta t=0.01~\text{fs}$.
We measure the errors of spatial marginal distribution of the Wigner function, 
\begin{equation}
\label{eq:fsm}
F_{sm}(x_1,x_2, t) = \iint_{\mathcal{K}\times\mathcal{K}} f(x_1, x_2, k_1, k_2, t) \, \dif k_1 \dif k_2, 
\end{equation}
in a similar way to calculating Eqs.~\eqref{l2} and \eqref{lf}. Fig.~\ref{fig:2dDirac_error} presents the errors against $N_k$ and $M$ after fixing $N_{um}=400$ in Eq.~\eqref{Mesh_um} and the spectral convergence is evident again. 
The spatial marginal distributions on the finest mesh at different instants are displayed 
in Fig.~\ref{fig:delta_fermidirac82d} where the corresponding constant distributions in the free space, $F_{sm}^{free}(x_1,x_2, t) \equiv 0.05384$,  have been subtracted. It can be observed there that the Fermi-Dirac distribution first reacts strongly to the eight singular points in the circle during which many small oscillations are produced in the central area of the circle, and then gradually expands to the surroundings. Obviously, such expanding is blocked by the eight Dirac delta function potentials and the resulting interference forms $12$ branches outside the circle:
$4$ big branches lie in the main directions, north, east, south and west, respectively, and the remaining $8$ small branches are equally distributed between them,
where six lines that are determined by pairs of singular points: $(1, 7)$, $(2,6)$, $(3,5)$,
$(1,3)$, $(4,8)$ and $(5,7)$ serve as the boundaries of branches. Inside the circle, the interference pattern shows a clear square structure that is also shaped by the same six lines. At the final time $t=2.5 ~\text{fs} $, 
a basin structure emerges: The spatial marginal distribution inside the circle is reduced to less than $F_{sm}^{free}$ (see Fig.~\ref{fig:delta_fermidirac8fd2d_t25_contour}), and that above $F_{sm}^{free}$ is all outside the circle (see Fig.~\ref{fig:delta_fermidirac8fd2d_t25}).

\section{Extensions to other singular potentials}
\label{sec:exte}

In this section, we devote ourselves into using the Wigner function approach to the following singular potentials: 
\begin{itemize}
\item The logarithmic potential
\begin{equation}
\label{V_log}
 V(x) = H \log (x)
\end{equation}
which is naturally related to the entropy expression \cite{gal2017nonlocal};

\item The inverse power potential for $\alpha\in (0,1)$
\begin{equation}
\label{V_power}
V (x) = H |x|^{-\alpha}
\end{equation}
which can be found in various quantum mechanical models \cite{meetz1964singular, serber1964scaling, tiktopoulos1965high};

\item The inverse square  potential
\begin{equation}
\label{V_fractional}
V (x) = H |x|^{-2}
\end{equation}
which has strong singularity at $x=0$ and is extensively used in high-energy scattering studies \cite{tiktopoulos1965high}.
\end{itemize}

\subsection{The logarithmic potential}
\label{sec:log}

\begin{figure}[htbp!]
	\centering 
\subfigure
{\includegraphics[width=0.495\textwidth,height=0.38\textwidth]{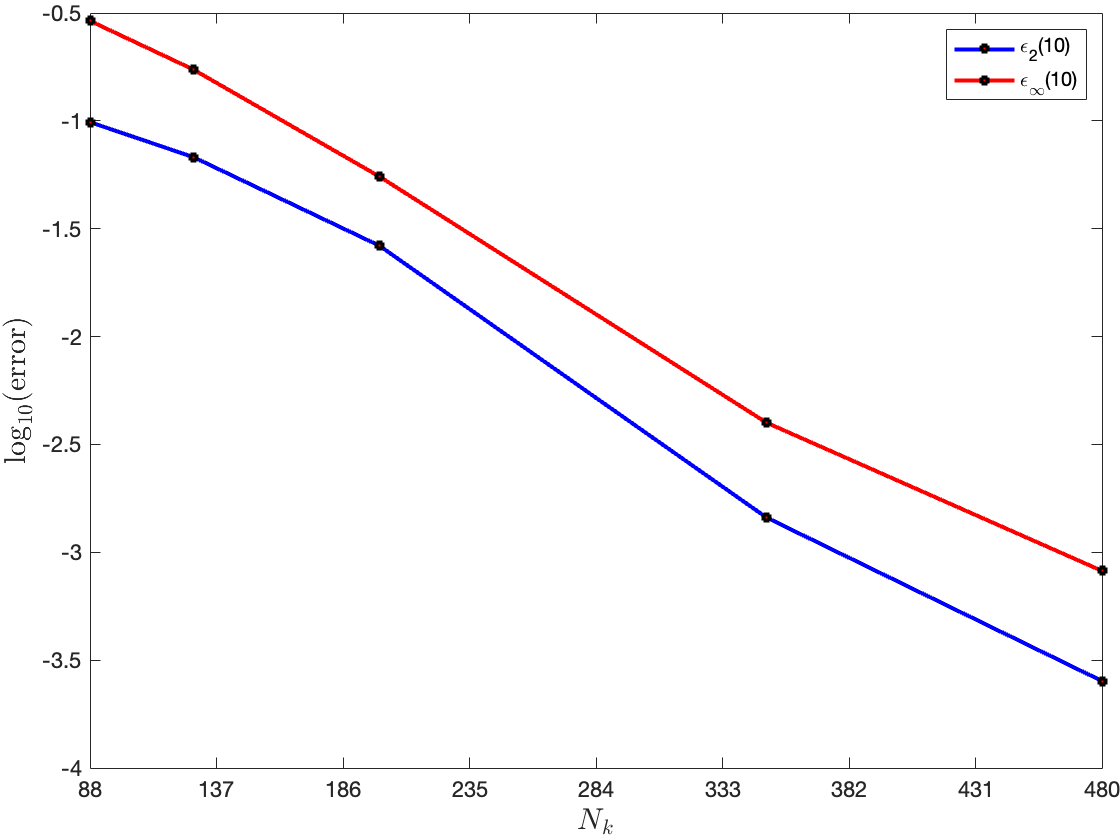}}
\subfigure
{\includegraphics[width=0.495\textwidth,height=0.38\textwidth]{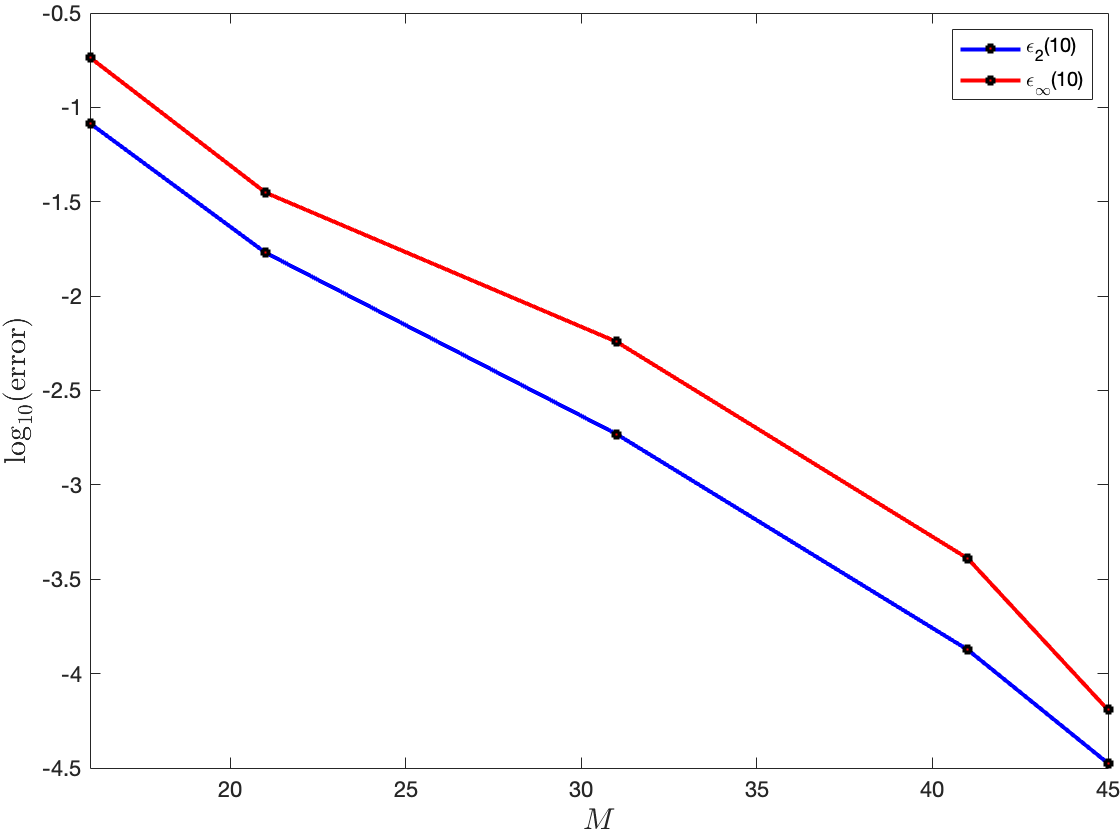}}
\caption[Convergence order log]{\small The logarithmic potential: Numerical convergence against $N_k$ (left) and  $M$ (right) at $t_{fin}=10$~fs.} 
	\label{fig:log_error} 
\end{figure}

\begin{figure}[htbp!]
	\centering 
\subfigure[ $t=2.5 ~\text{fs} $.]
{\includegraphics[width=0.326\textwidth,height=0.23\textwidth]{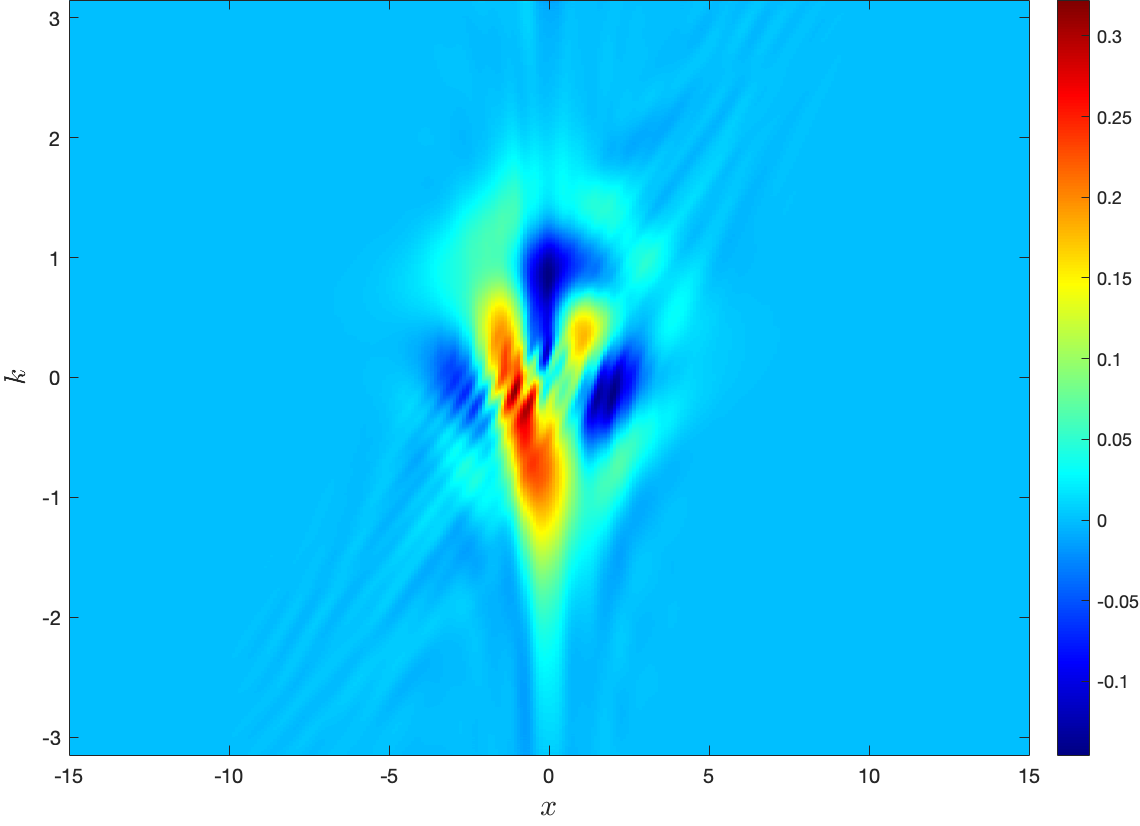}
\includegraphics[width=0.326\textwidth,height=0.23\textwidth]{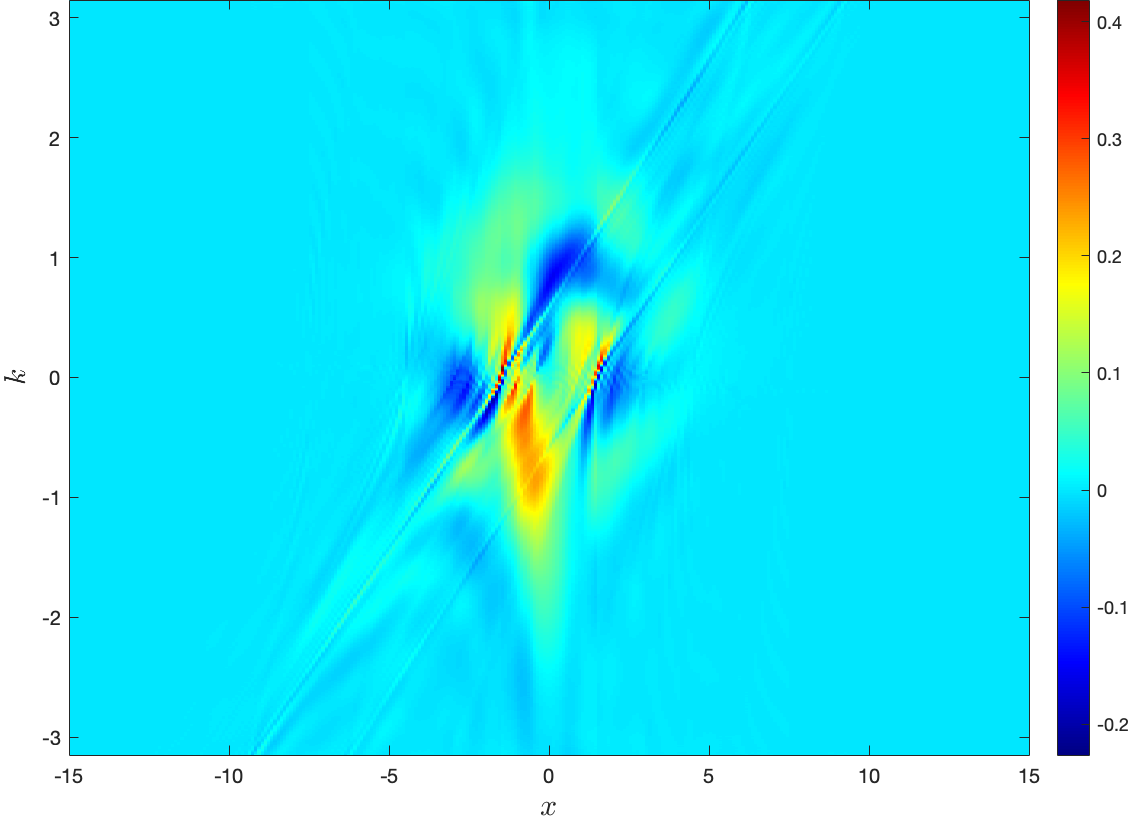}
\includegraphics[width=0.326\textwidth,height=0.23\textwidth]{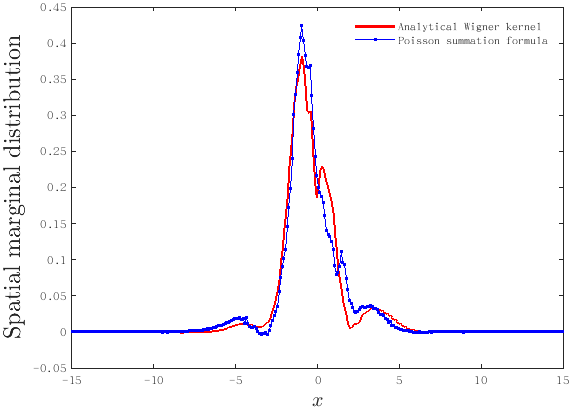}}\\
\subfigure[ $t=5 ~\text{fs} $.]
{\includegraphics[width=0.326\textwidth,height=0.23\textwidth]{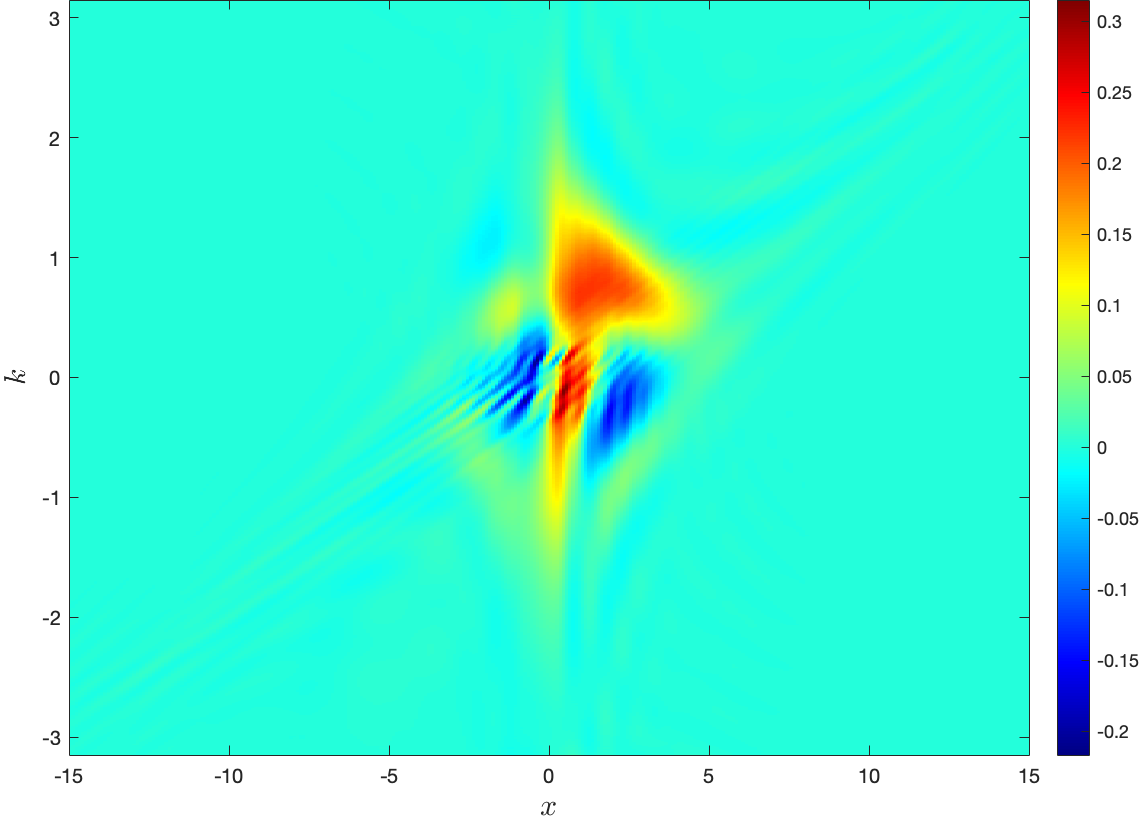}
\includegraphics[width=0.326\textwidth,height=0.23\textwidth]{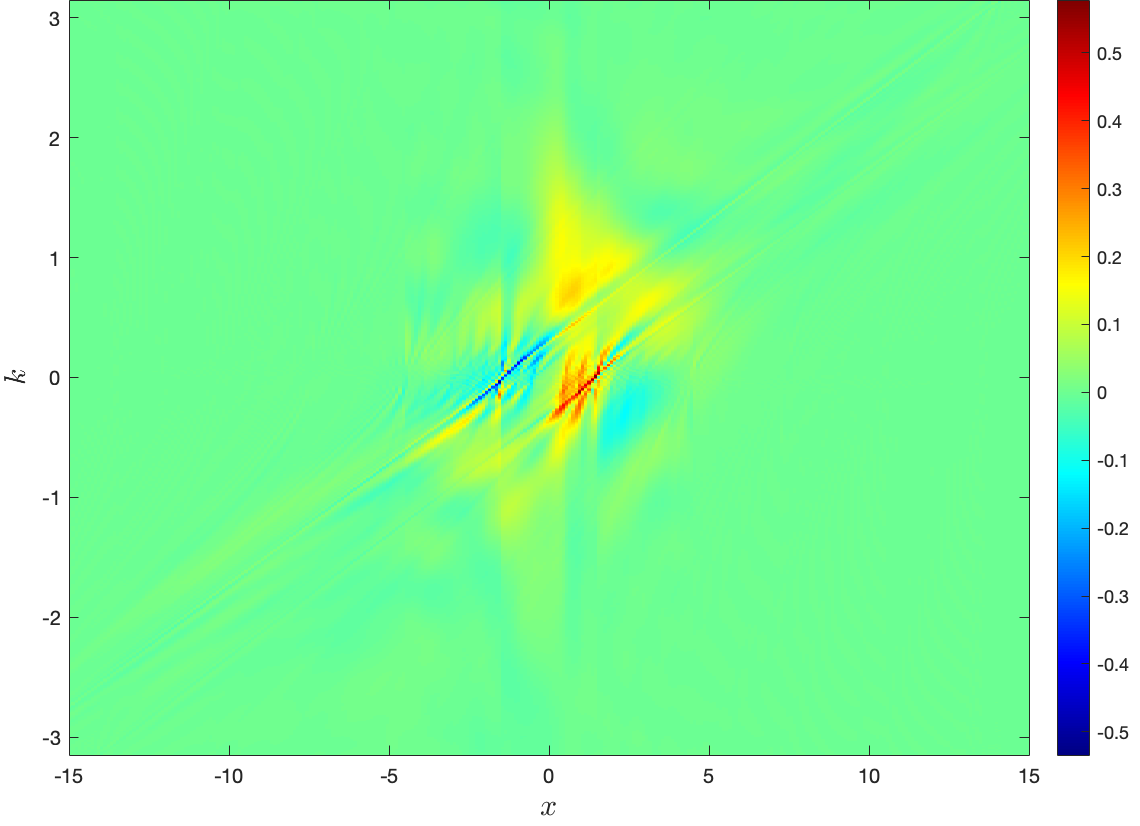}
\includegraphics[width=0.326\textwidth,height=0.23\textwidth]{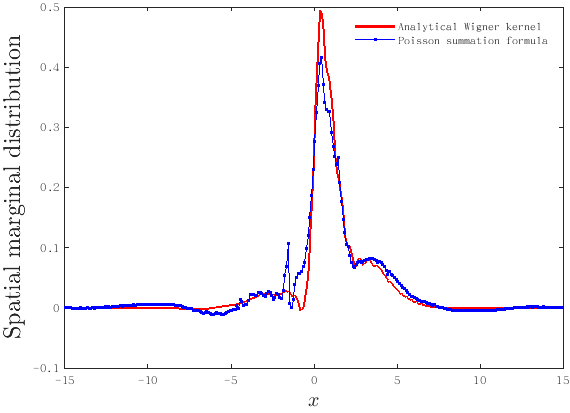}}\\
\subfigure[ $t=7.5 ~\text{fs} $.]
{\includegraphics[width=0.326\textwidth,height=0.23\textwidth]{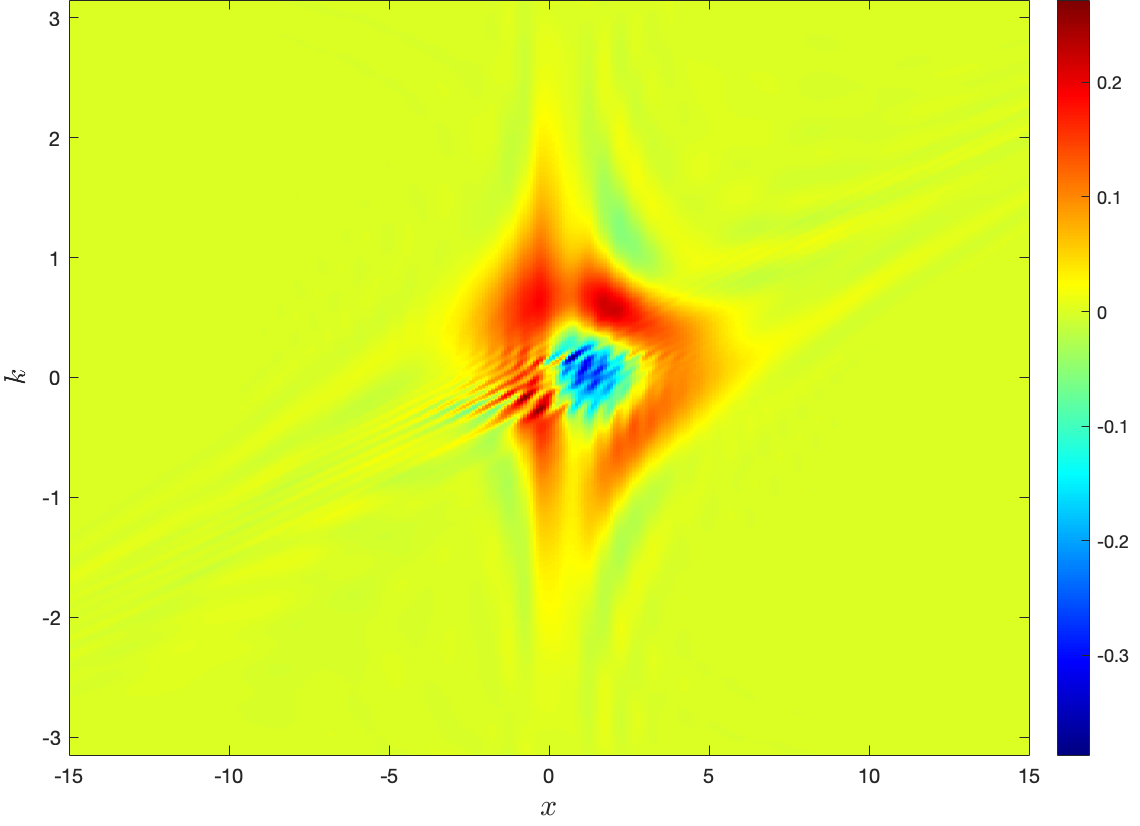}
\includegraphics[width=0.326\textwidth,height=0.23\textwidth]{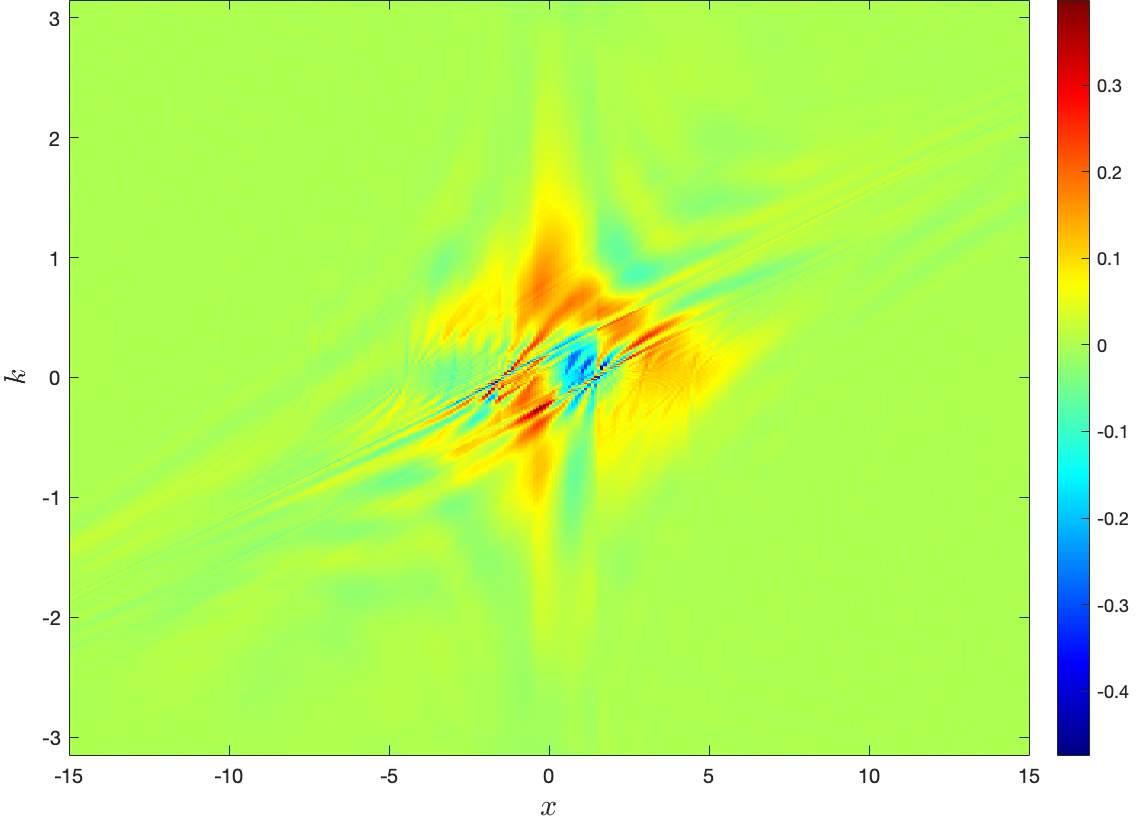}
\includegraphics[width=0.326\textwidth,height=0.23\textwidth]{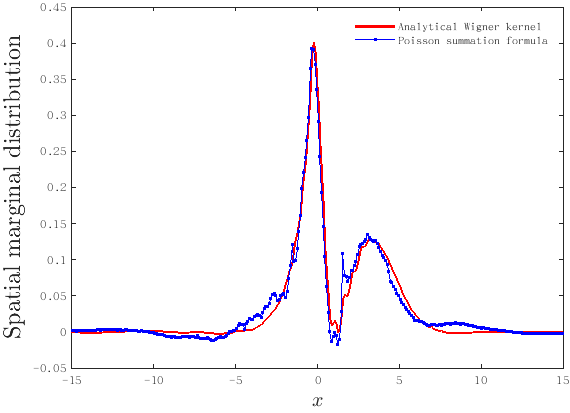}}\\
\subfigure[ $t=10 ~\text{fs} $.]
{\includegraphics[width=0.326\textwidth,height=0.23\textwidth]{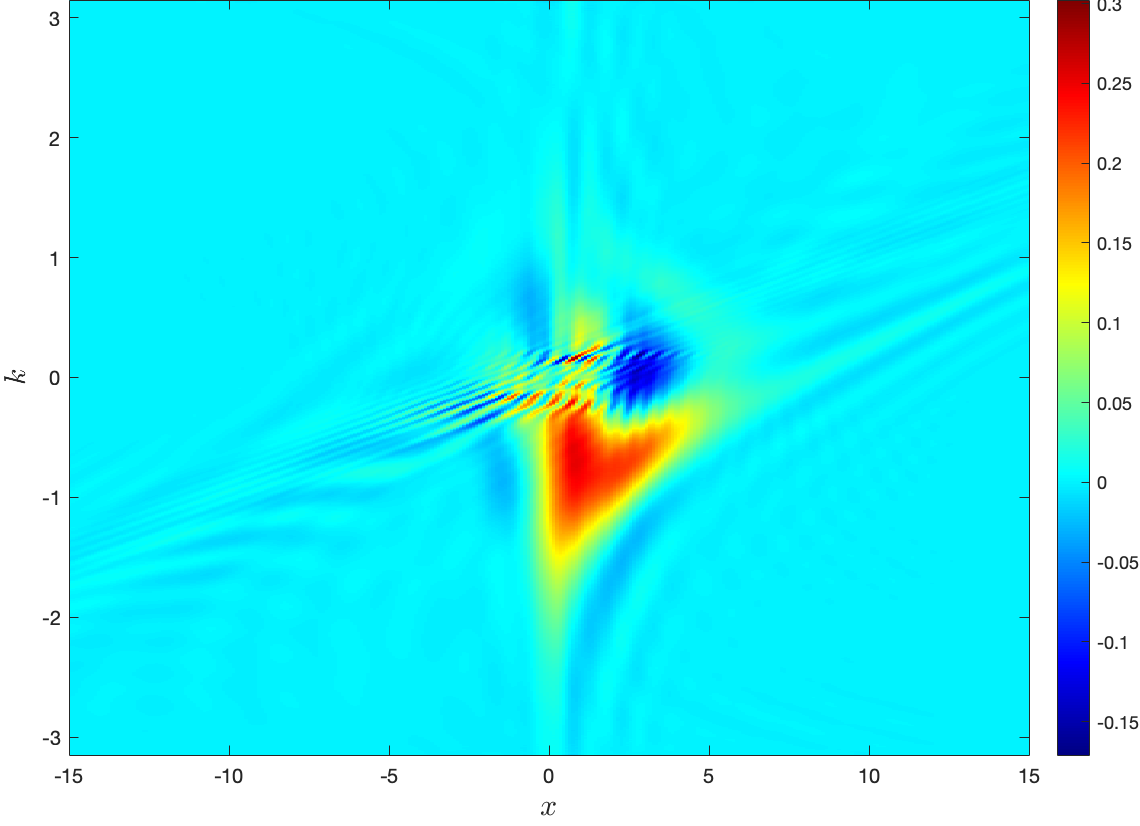}
\includegraphics[width=0.326\textwidth,height=0.23\textwidth]{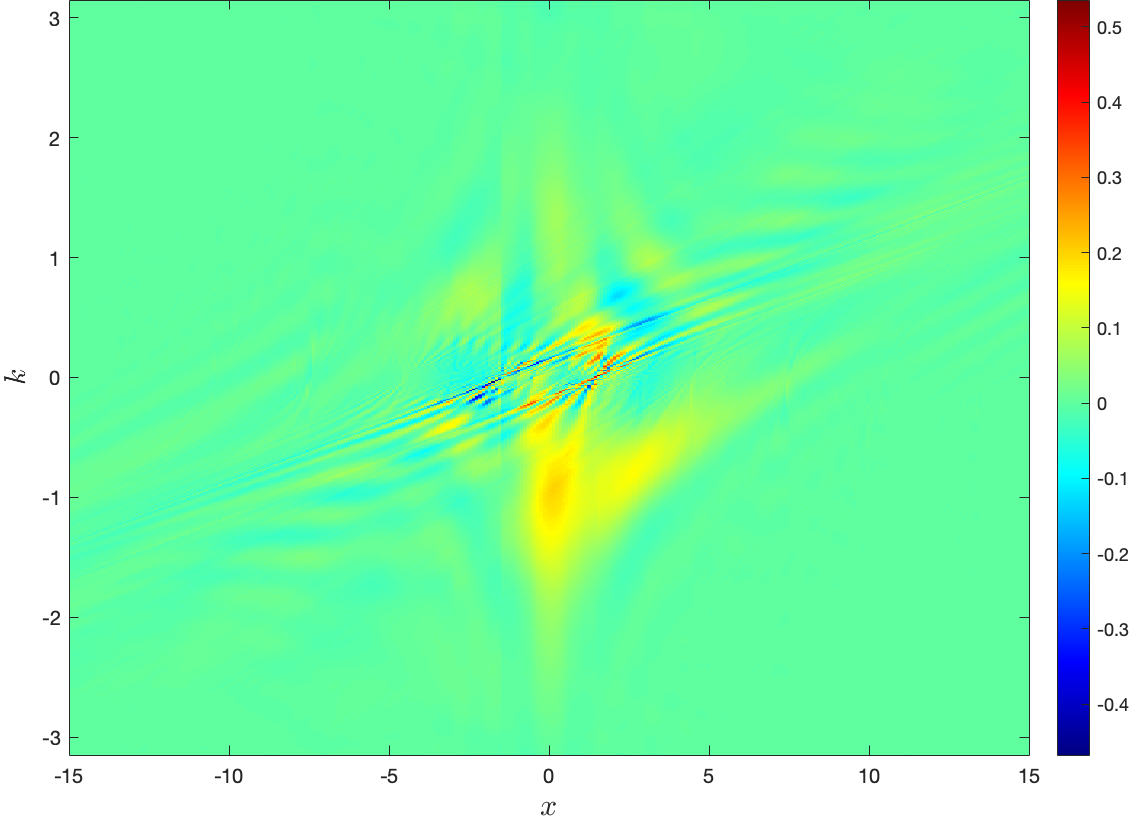}
\includegraphics[width=0.326\textwidth,height=0.23\textwidth]{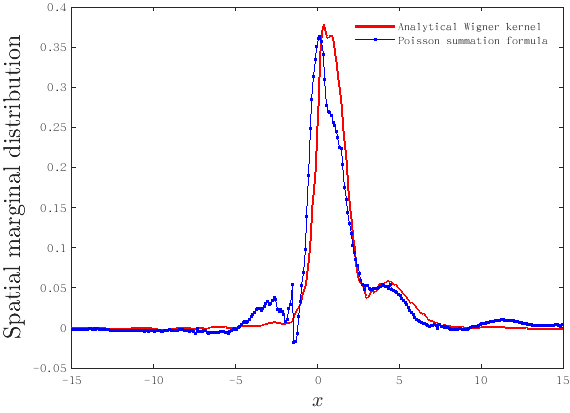}}\\
\caption[Log1.]{\small The logarithmic potential: 
The Wigner functions corresponding to the analytical Wigner kernel \eqref{Vw_log} (left),
to the ``approximate" Wigner kernel \eqref{poi} (middle) and 
their spatial marginal distributions (right).} 
\label{fig:log1} 
\end{figure}

Plugging Eq.~\eqref{V_log} into Eq.~\eqref{Vw} yields the corresponding Wigner kernel
\begin{equation}
\label{Vw_log}
V_w(x,k) = -\frac{H}{\hbar}\frac{\sin(2xk)}{|k|},
\end{equation}
and then substituting it in Eq.~\eqref{c} leads to
\begin{equation}
\label{c_log}
 \frac{\hbar}{2H\mi} c_\nu( x) =  \int_{0}^{L_k} \frac{\sin(2xk^\prime) \sin(\tilde{\nu} k^\prime )}{k^\prime}    \dif  k^\prime
 : =  \left( \int_0^\varepsilon +\int_\varepsilon^{L_k} \right) g_{\nu}(x, k^\prime) \dif  k^\prime,
\end{equation}
where $\varepsilon$ is a prescribed small parameter and $\tilde{\nu} = 2\pi \nu/L_k$. Using the the Taylor expansion in $(0, \varepsilon)$ gives
\[
\int_0^\varepsilon g_{\nu}(x, k^\prime) \dif  k^\prime = \tilde{\nu} x \varepsilon^2 - (\frac13 \tilde{\nu} x^3 + \frac{1}{12} \tilde{\nu}^3 x)\varepsilon^4 + \mathcal{O}(\varepsilon^7),
\]
and with the help of the cosine integral function $\operatorname{Ci} (x) = - \int_x^{+\infty}  \frac{\cos t}{t}\, \dif t$,
we have
\[
\int_\varepsilon^{L_k}  g_{\nu}(x, k^\prime) \dif  k^\prime = \frac{\operatorname{Ci} (|\omega_\nu^+(x)| \varepsilon)-\operatorname{Ci} (|\omega_\nu^-(x)| \varepsilon)-\operatorname{Ci} (|\omega_\nu^+(x)| {L_k})+\operatorname{Ci} (|\omega_\nu^-(x)| {L_k})}{2}.
\]
Accordingly,  the expansion coefficient $c_\nu( x)$ given in an oscillatory improper integral \eqref{c_log}
can be approximated to the machine resolution by choosing $\varepsilon = 1$E-5.
Other parameters are set to be:  $H=1$, $-X_L=X_R=30 ~\text{nm}$, $-k_{min}=k_{max}=\pi ~\text{nm}^{-1}$,
$N_k=512$, $Q=20$, $M=55$, $\Delta t=0.01~\text{fs}$, and $N_{um}=600$. 
We use an initial Gaussian wave packet close to the origin by setting $x^0 =-1 ~\text{nm}$ and $k^0=0.5 ~\text{nm}^{-1}$ 
in Eq.~\eqref{GaussF}. Numerical convergence tests are given in Fig.~\ref{fig:log_error} and show clearly the spectral accuracy. The left column of Fig.~\ref{fig:log1} plots the numerically converged Wigner functions at $t=2.5,\, 5,\, 7.5, \,10$~fs obtained on the finest mesh. It can be observed there that, the wave packet is attracted by the logarithmic potential \eqref{V_log},
and keeps moving around the singular point, during which many small oscillations appear around the origin along with the singularity.

It has been shown that the Poisson summation formula can be used to approximate the Wigner kernel \eqref{Vw} well ($y_\zeta = \zeta \,\Delta y$ and $\Delta y$ being the spacing):
\begin{equation}
\label{poi}
V_w (x,k) \approx  \frac{\Delta y}{2\pi \mi \hbar} \sum_{\zeta=-\infty}^{+\infty} \left[ V(  x+\frac{   y_\zeta}{2})-V(  x-\frac{   y_\zeta}{2}) \right] \, e^{-\mi  k y_\zeta}
\end{equation}
when the external potential $V(x)$ is smooth and localized \cite{xiong2016advective}, 
but fails when taking the Coulomb interaction into account \cite{xiong2022characteristic}. 
Here we would like to confirm this failure using the logarithmic potential. 
After using the ``approximate" Wigner kernel \eqref{poi} to replace the analytical one \eqref{Vw_log} 
and keeping all other settings unchanged, we rerun the simulations in the left column of Fig.~\ref{fig:log1} 
and display the resulting Wigner functions in the middle column of Fig.~\ref{fig:log1},
which shows some obvious discrepancy. 
Compared with the reference solutions, the Wigner functions obtained with Eq.~\eqref{poi} 
display much more severe oscillations with higher peaks and deeper valleys,
and the corresponding spatial marginal distributions show 
some spurious oscillations and non-physical negative values (see the right column of Fig.~\ref{fig:log1}).

\subsection{The inverse power  potential}
\label{sec:power}
\begin{figure}[htbp!]
\centering 
\subfigure
{\includegraphics[width=0.495\textwidth,height=0.38\textwidth]{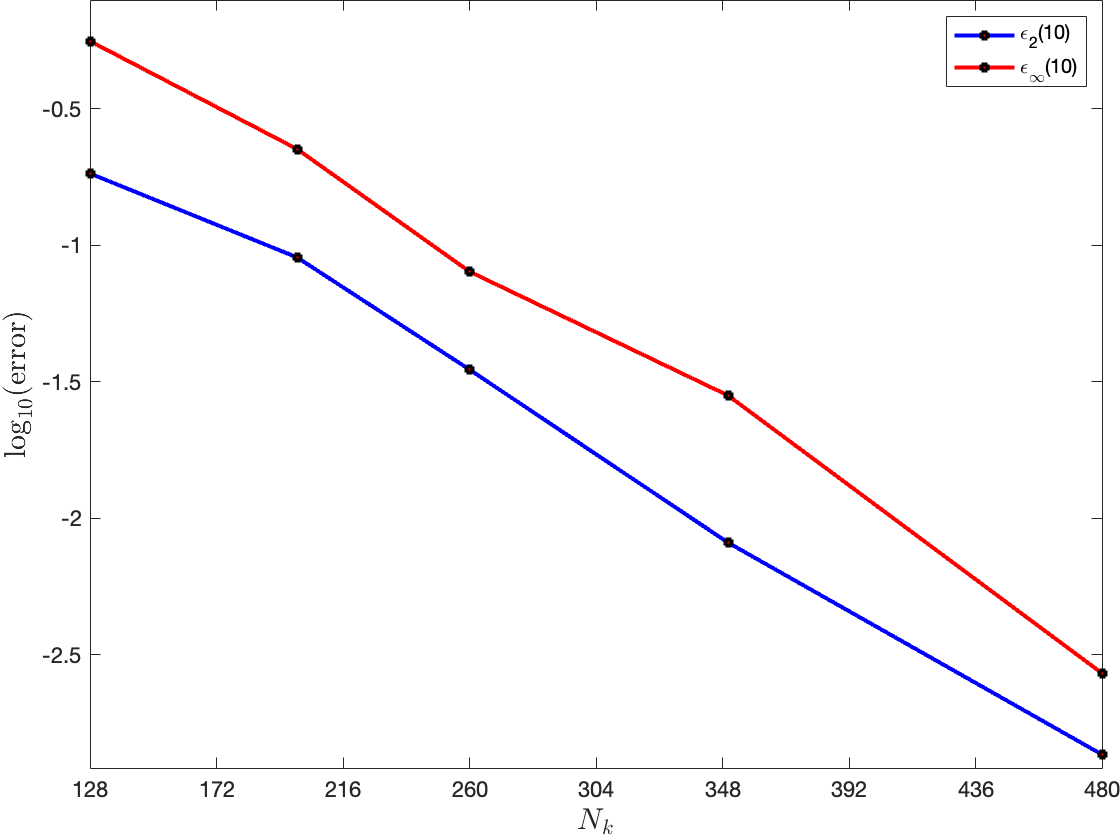}}
\subfigure
{\includegraphics[width=0.495\textwidth,height=0.38\textwidth]{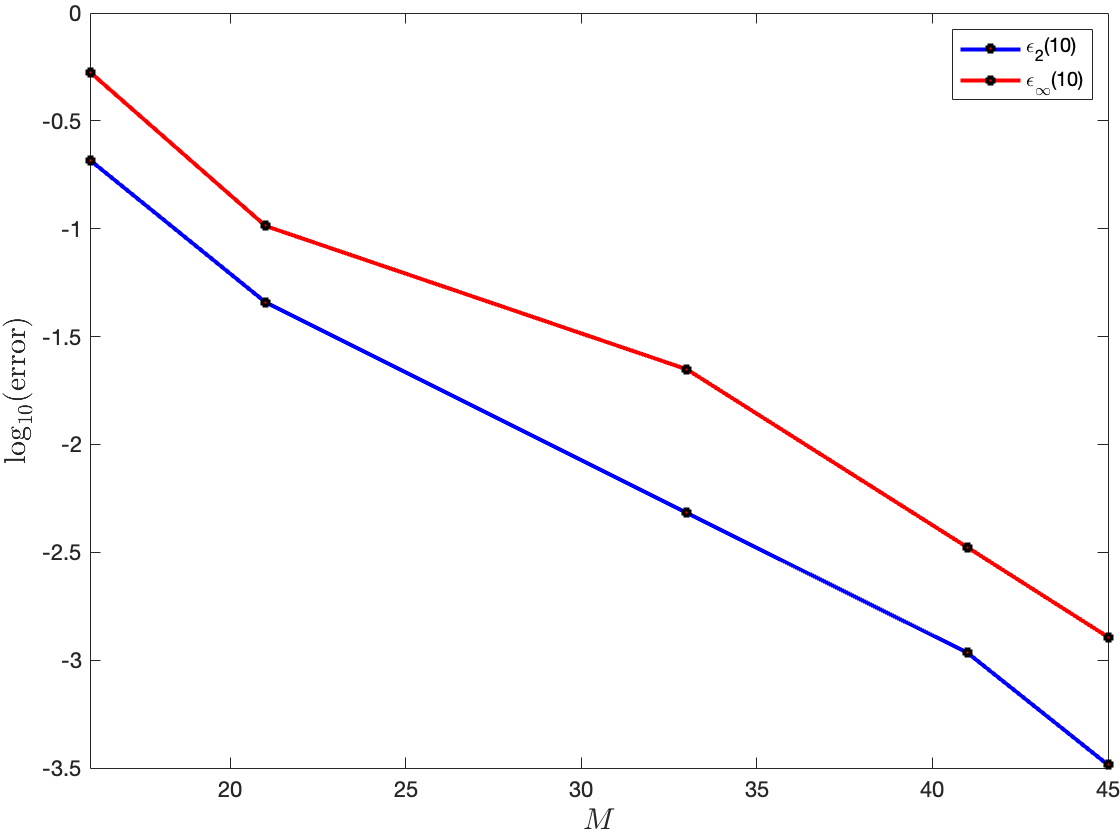}}
\caption[Convergence order power]{\small The inverse power potential: Numerical convergence with respect to $N_k$ (left) and  $M$ (right) at $t_{fin}=10$~fs.} 
\label{fig:power_error} 
\end{figure}

\begin{figure}[htbp!]
\centering 
\subfigure[ $t=2.5$~fs.]
{\includegraphics[width=0.495\textwidth,height=0.38\textwidth]{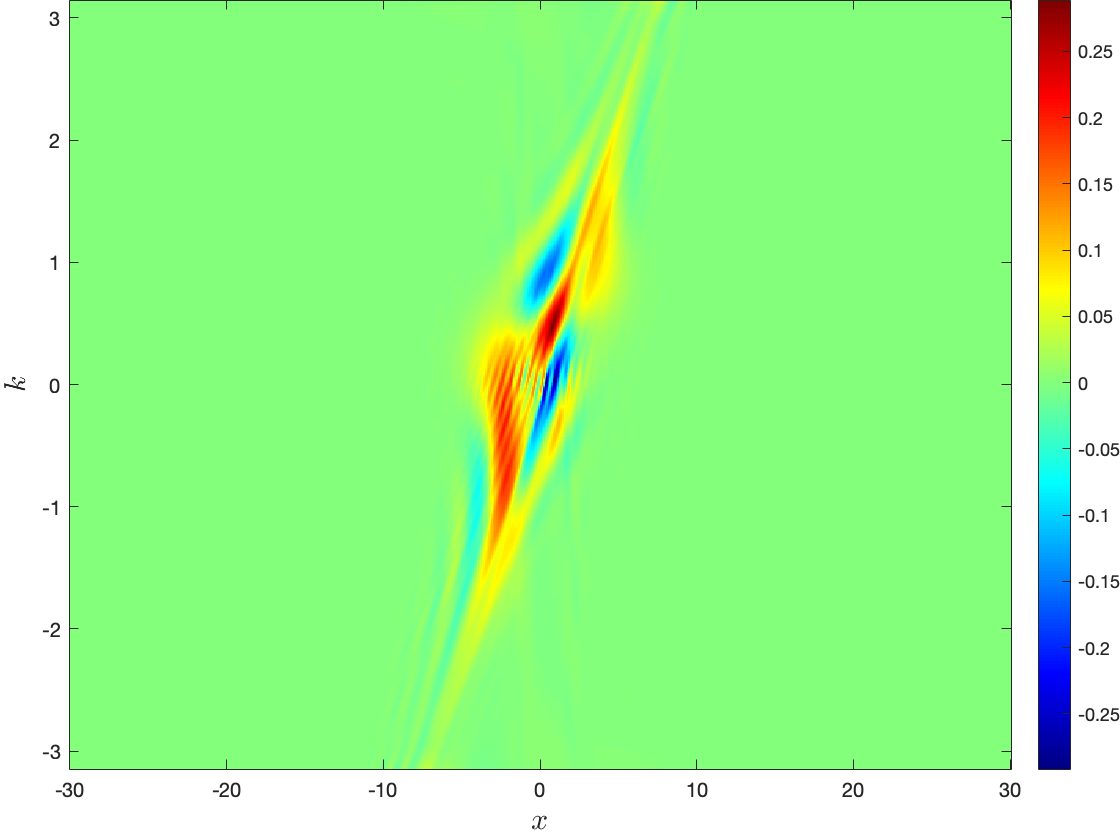}}
\subfigure[ $t=5$~fs.]
{\includegraphics[width=0.495\textwidth,height=0.38\textwidth]{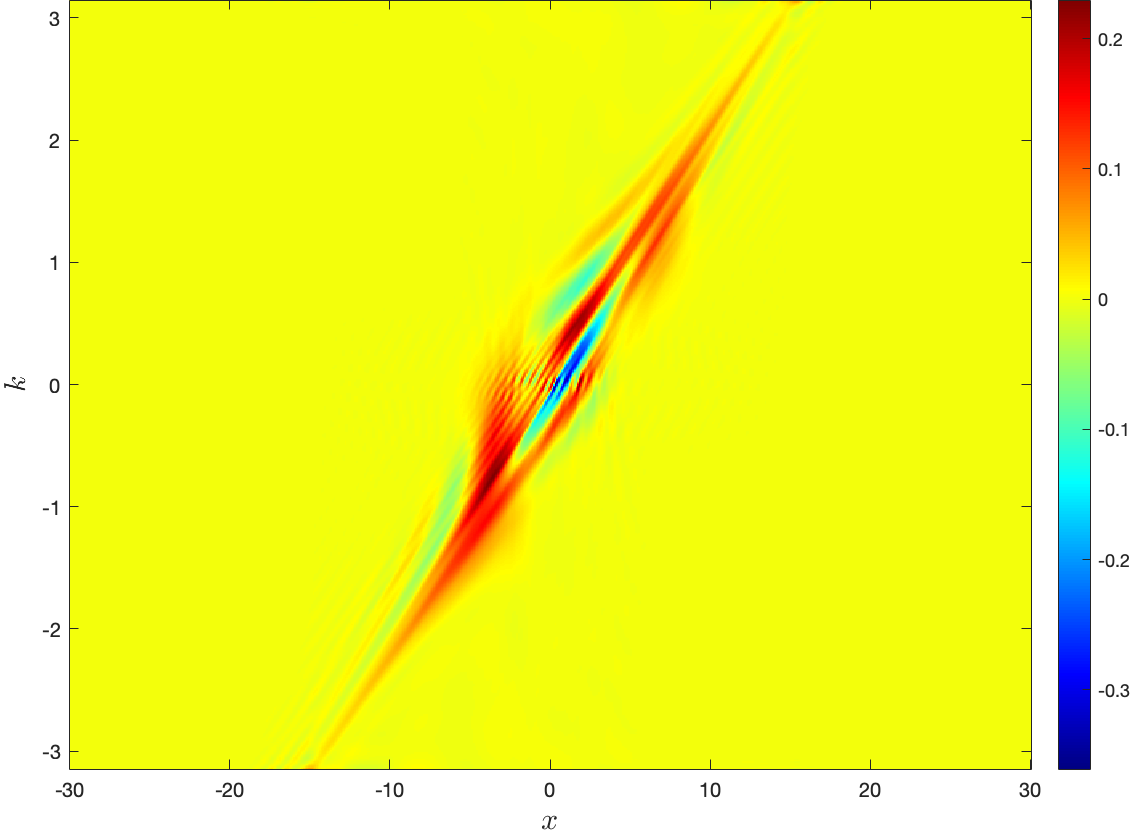}}\\
\subfigure[ $t=7.5$~fs.]
{\includegraphics[width=0.495\textwidth,height=0.38\textwidth]{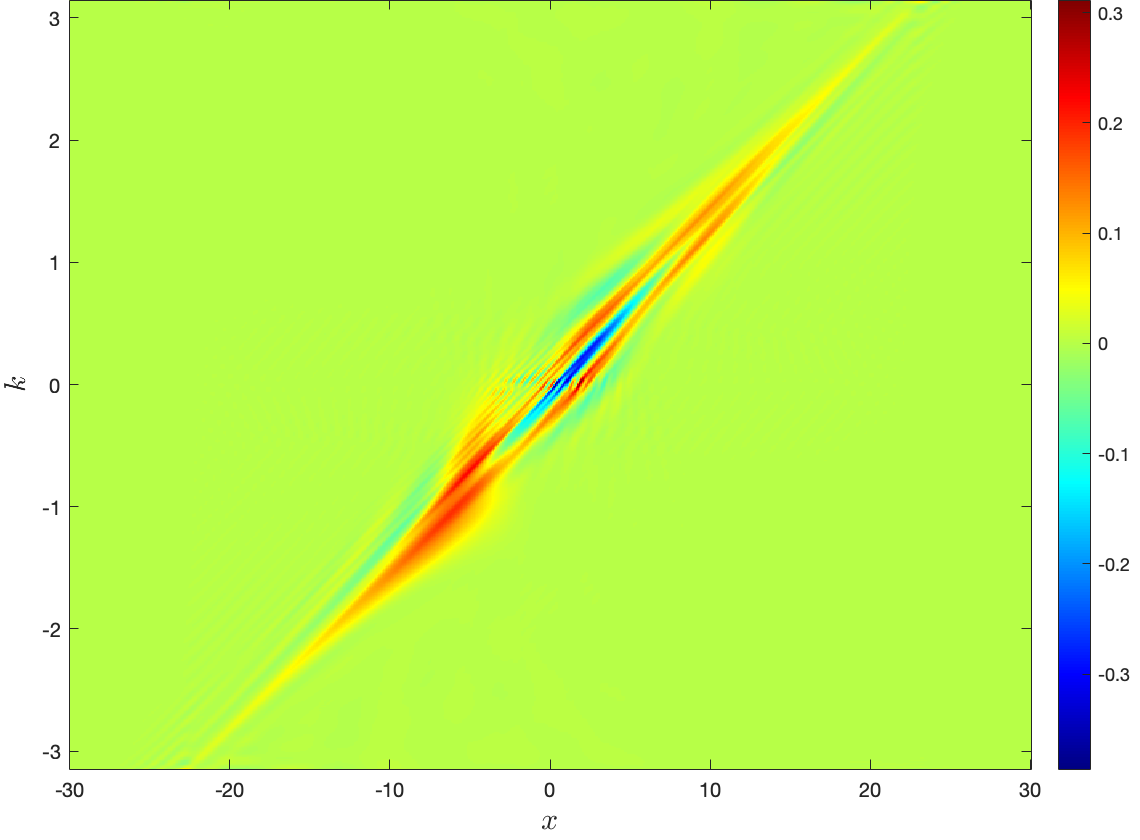}}
\subfigure[ $t=10$~fs.]
{\includegraphics[width=0.495\textwidth,height=0.38\textwidth]{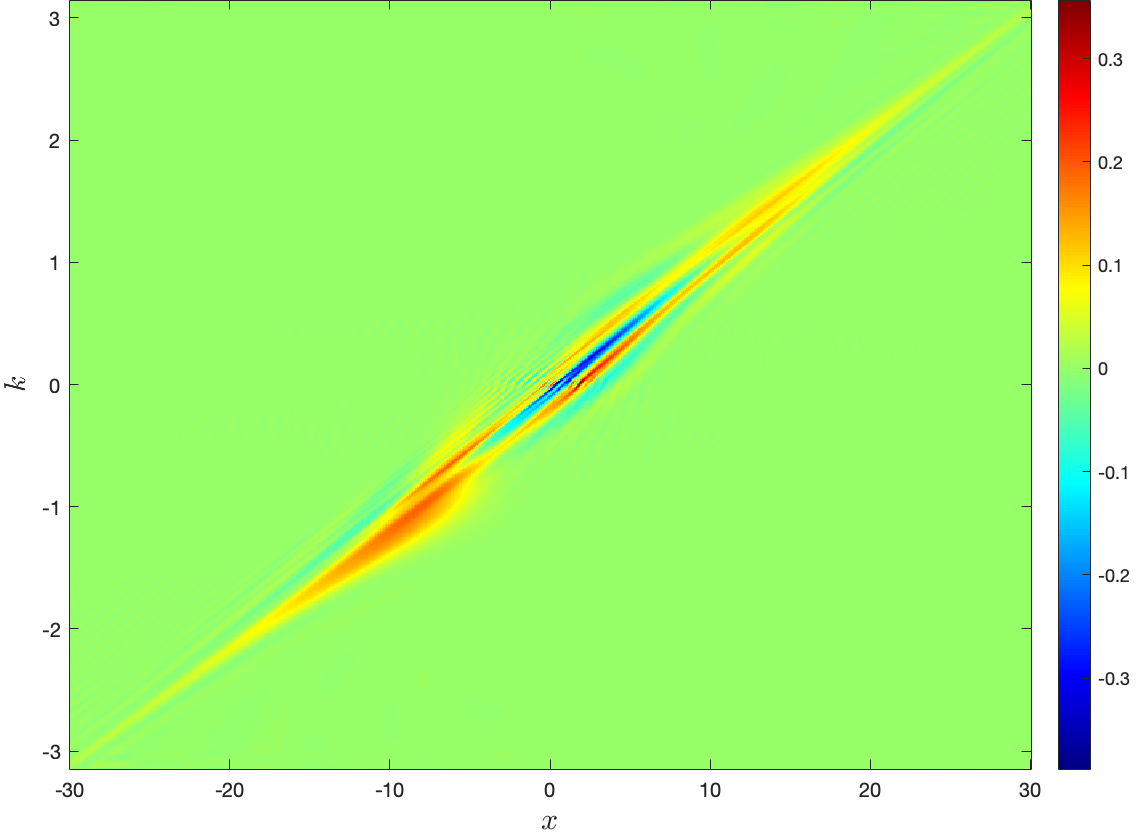}}\\
\caption[Power.]{\small The inverse power potential: The Wigner functions at different instants.  We set $H=1$ and
$\alpha=1/2$ in Eq.~\eqref{V_power}.} 
\label{fig:power} 
\end{figure}

\begin{figure}[htbp!]
\centering 
\subfigure[Tunneling rate.]
{\includegraphics[width=0.495\textwidth,height=0.38\textwidth]{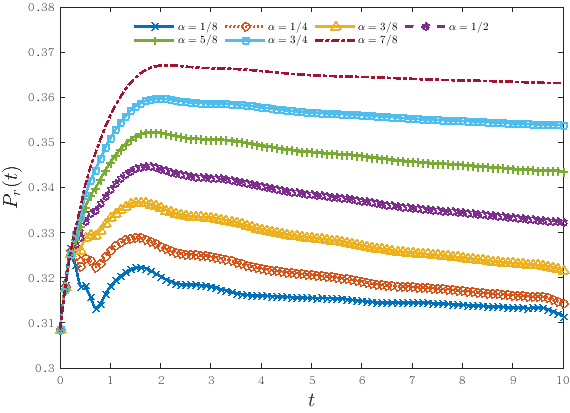}}
\subfigure[Uncertainty.]
{\includegraphics[width=0.495\textwidth,height=0.38\textwidth]{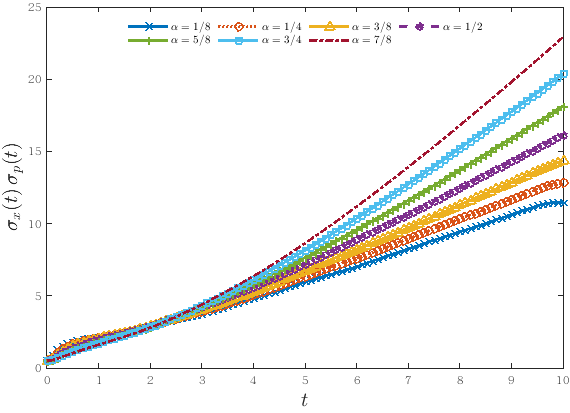}}
\caption[Compower\_Power.]{ \small The inverse power potential: Tunneling rates and uncertainties. It is clearly seen that the uncertainty $\sigma_x (t)\, \sigma_p(t)$ shows a mounting tendency whilst $P_r(t)$ ascends.} 
\label{fig:phy_power} 
\end{figure}

According to the Fourier transform of the inverse power function with $\alpha \in (0,1)$: $\mathcal{F} [|x|^\alpha] (\xi) =  -2 \sin(\frac{\pi}{2}(\alpha-1)) \Gamma(\alpha)  |\xi|^{-\alpha}$, the Wigner kernel of Eq.~\eqref{V_power} reads 
\begin{equation}
\label{Vw_power}
V_w (x,k) = \frac{H \sin(\frac{\pi}{2}\alpha) \Gamma(\alpha) 2^{2-\alpha}}{\pi\hbar} \frac{\sin(2xk)}{|k|^{\alpha}},
\end{equation}
where $\Gamma(x)$ gives the Gamma function. Combining Eqs.~\eqref{Vw_power} and \eqref{c} together yields
\begin{equation}
c_\nu( x) = \frac{2H\mi \sin(\frac{\pi}{2}\alpha) \Gamma(\alpha) 2^{2-\alpha}}{\pi\hbar} \int_{0}^{L_k} \frac{\sin(2xk) \sin(2\pi \nu k /L_k)}{k^\alpha}  \,  \dif  k,  
\end{equation}
which can be efficiently approximated to the machine accuracy with the help of the Generalized Hypergeometric function ${}_{1}F_{2}((1-\alpha)/2; \,1/2, (3-\alpha)/2; \,x)$ \cite{whittaker1915course}. For example, when $\alpha = 1/2$, that Generalized Hypergeometric function reduces to the Fresnel function $\operatorname{C}(x) = \int_0^x \cos (\pi t^2 / 2) \, \dif t$. 
We adopt the same parameters as in Section~\ref{sec:log} to simulate the scattering between the Gauss wave packet and the inverse power potential. 
Fig.~\ref{fig:power_error} shows the spectral convergence with respect to $N_k$ and $M$ again,
and Fig.~\ref{fig:power} the Wigner functions at  four different instants. We are able to observe there that the wave packet partially passes through the inverse power potential barrier;
and the negative Wigner function,  sandwiched between two scattered wave packets in opposite directions, strongly implies the uncertainty principle around the singularity. Fig.~\ref{fig:phy_power} further plots the effect of power $\alpha$ on the tunneling rate $P_r(t)$ in Eq.~\eqref{Pr} and uncertainty $\sigma_x (t)\, \sigma_p (t)$ in Eq.~\eqref{uncertainty}.
It is evident that, the tunneling rate gradually increases as $\alpha$ rises, which reflects that the width of the potential becomes steadily smaller. 
Since $P_r(t)$ never rises over $0.5$ throughout the scattering, the uncertainty $\sigma_x (t)\, \sigma_p(t)$ shows a mounting tendency whilst $P_r(t)$ ascends, which is  incurred by the shape change of the potential, namely the increase of $\alpha$.

\subsection{The inverse square potential}

\begin{figure}[htbp!]
\centering 
\subfigure
{\includegraphics[width=0.495\textwidth,height=0.38\textwidth]{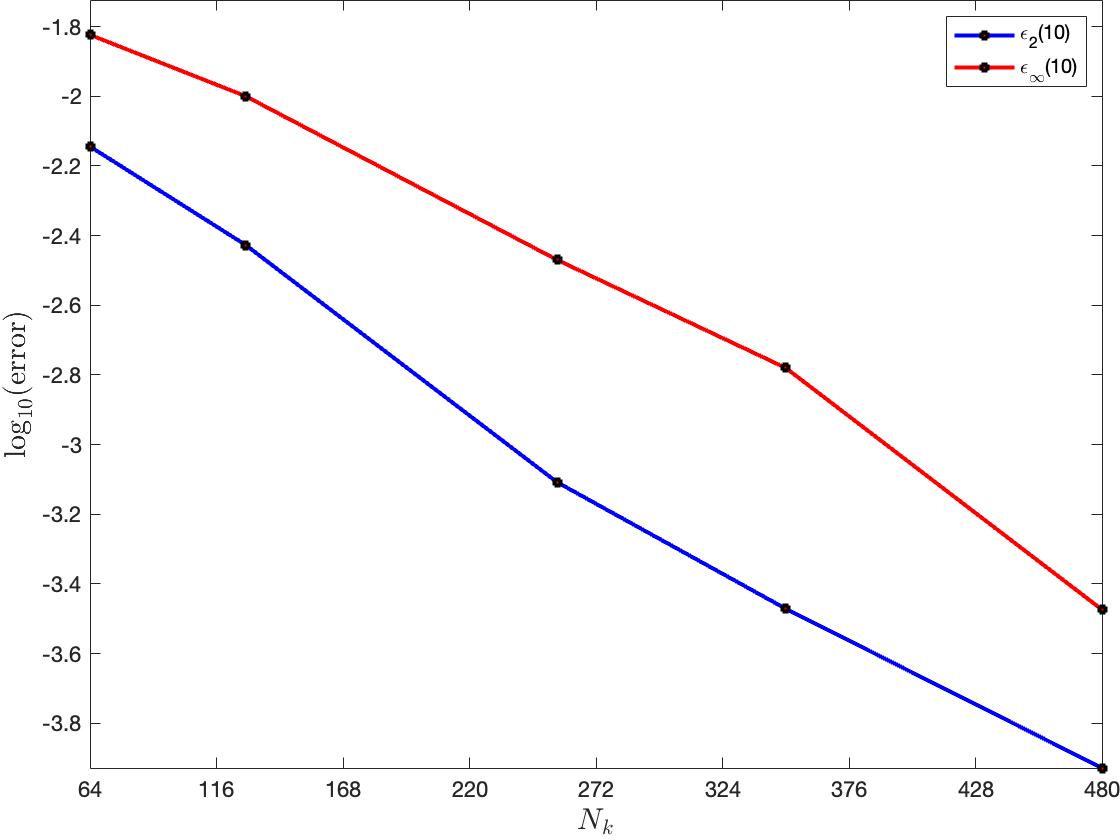}}
\subfigure
{\includegraphics[width=0.495\textwidth,height=0.38\textwidth]{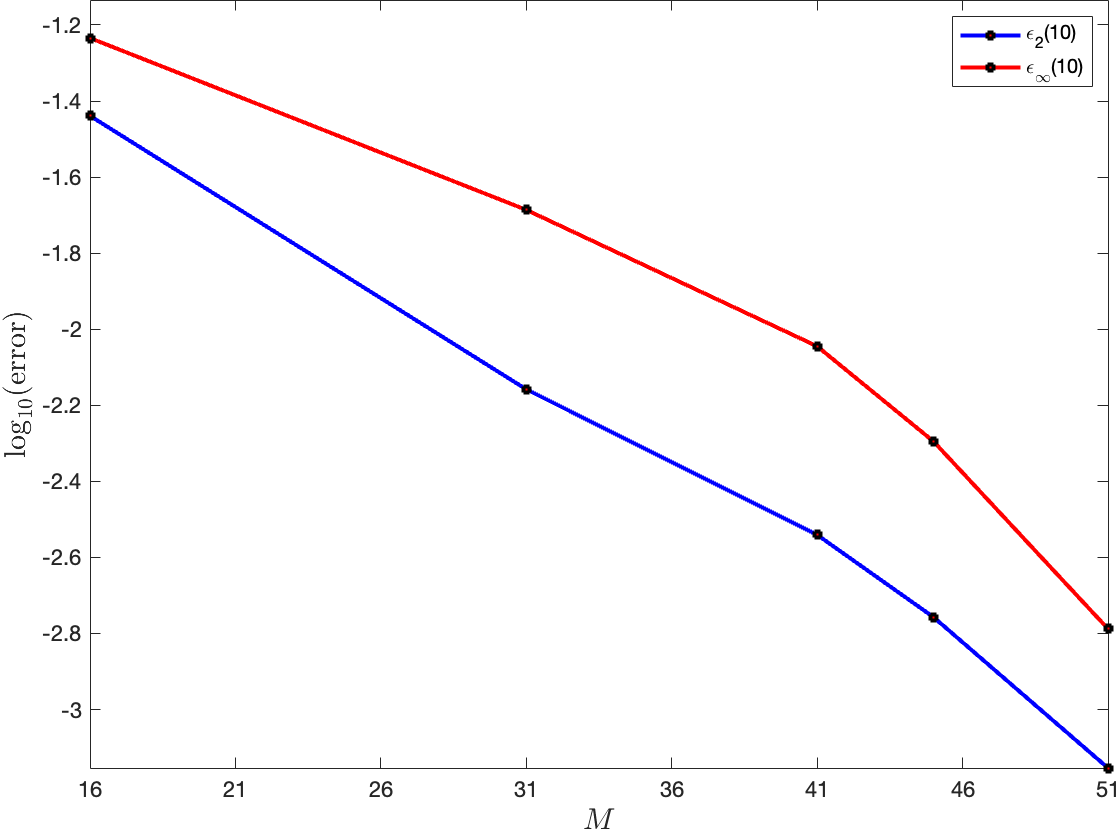}}
\caption[Convergence order inverse]{\small The inverse square potential: Spectral convergence against $N_k$ (left) and $M$ (right).} 
\label{fig:fractional_error} 
\end{figure}

\begin{figure}[htbp!]
\centering 
\subfigure[$t=2~\text{fs} $.]
{\includegraphics[width=0.495\textwidth,height=0.38\textwidth]{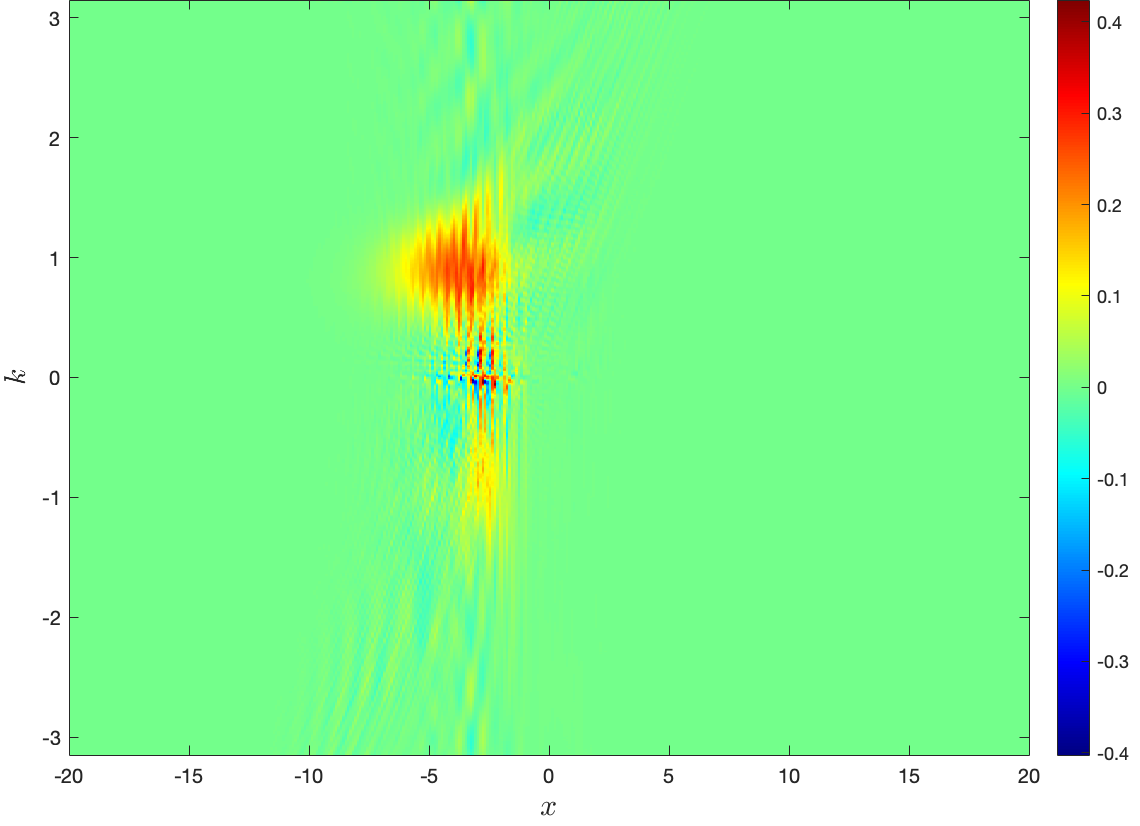}}
\subfigure[$t=4~\text{fs} $.]
{\includegraphics[width=0.495\textwidth,height=0.38\textwidth]{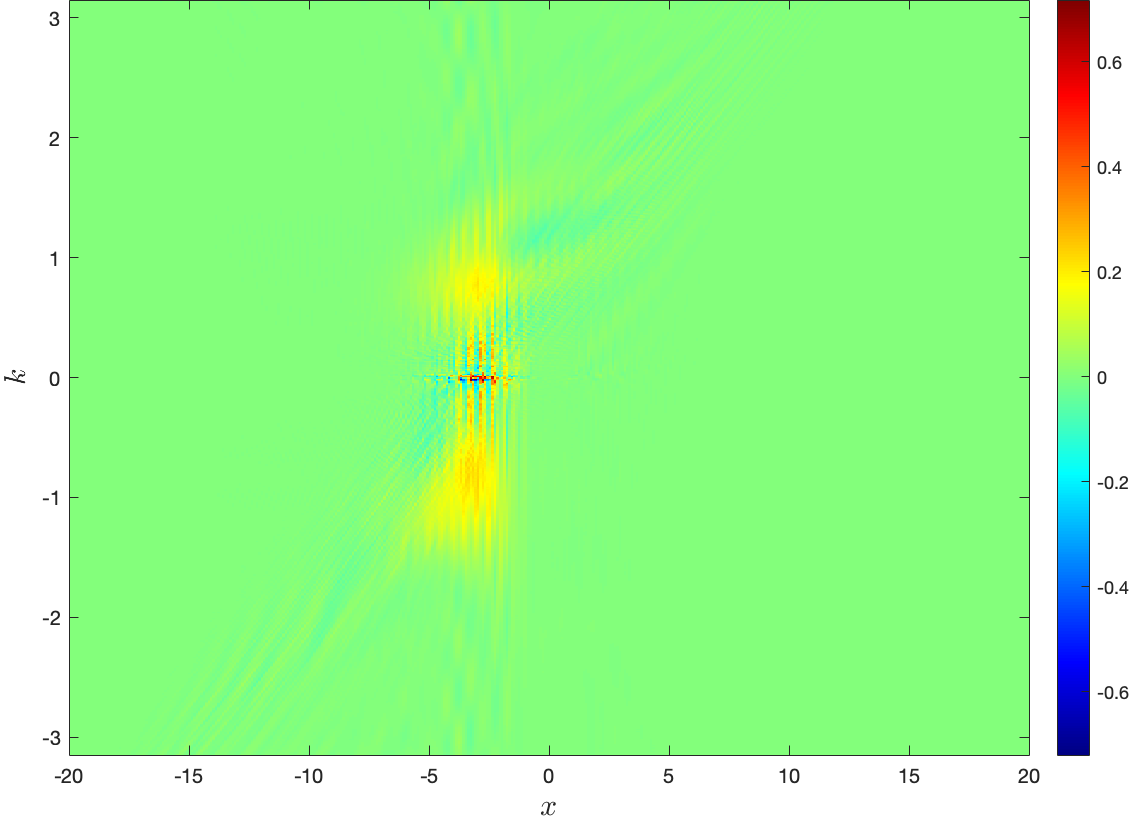}}\\
\subfigure[$t=5~\text{fs} $.]
{\includegraphics[width=0.495\textwidth,height=0.38\textwidth]{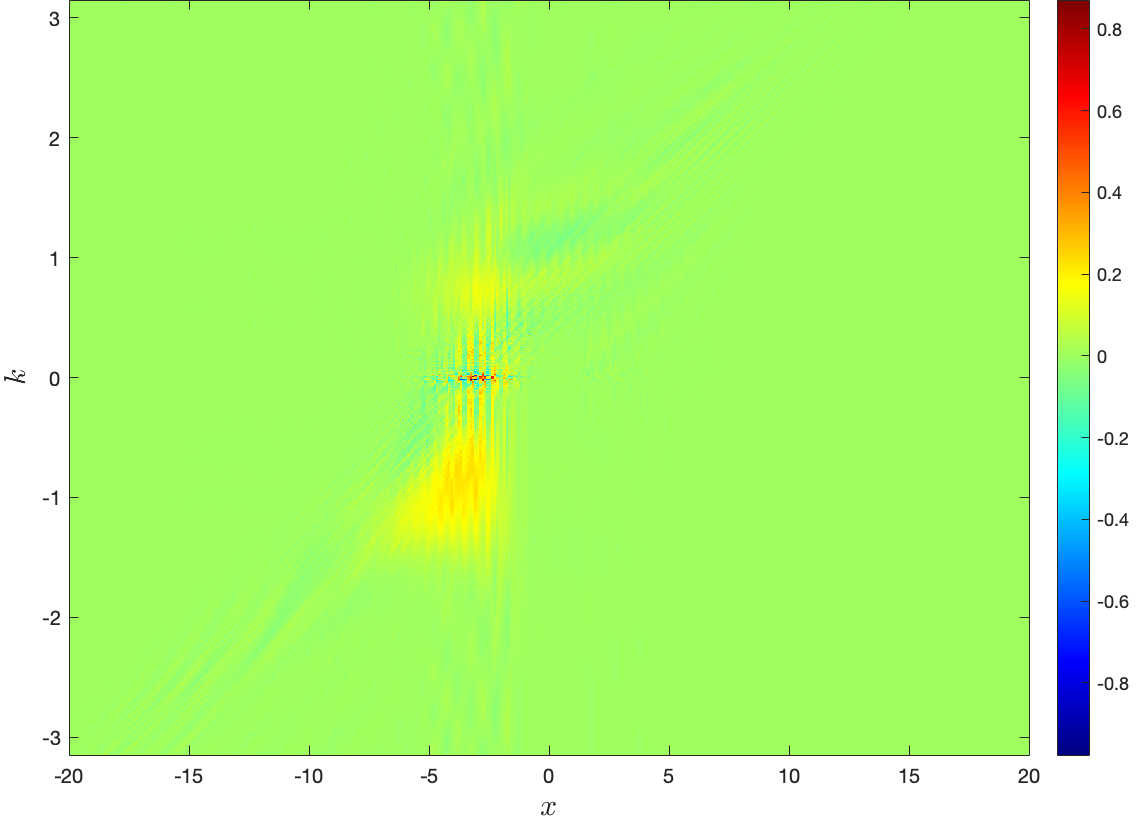}}
\subfigure[$t=8~\text{fs} $.]
{\includegraphics[width=0.495\textwidth,height=0.38\textwidth]{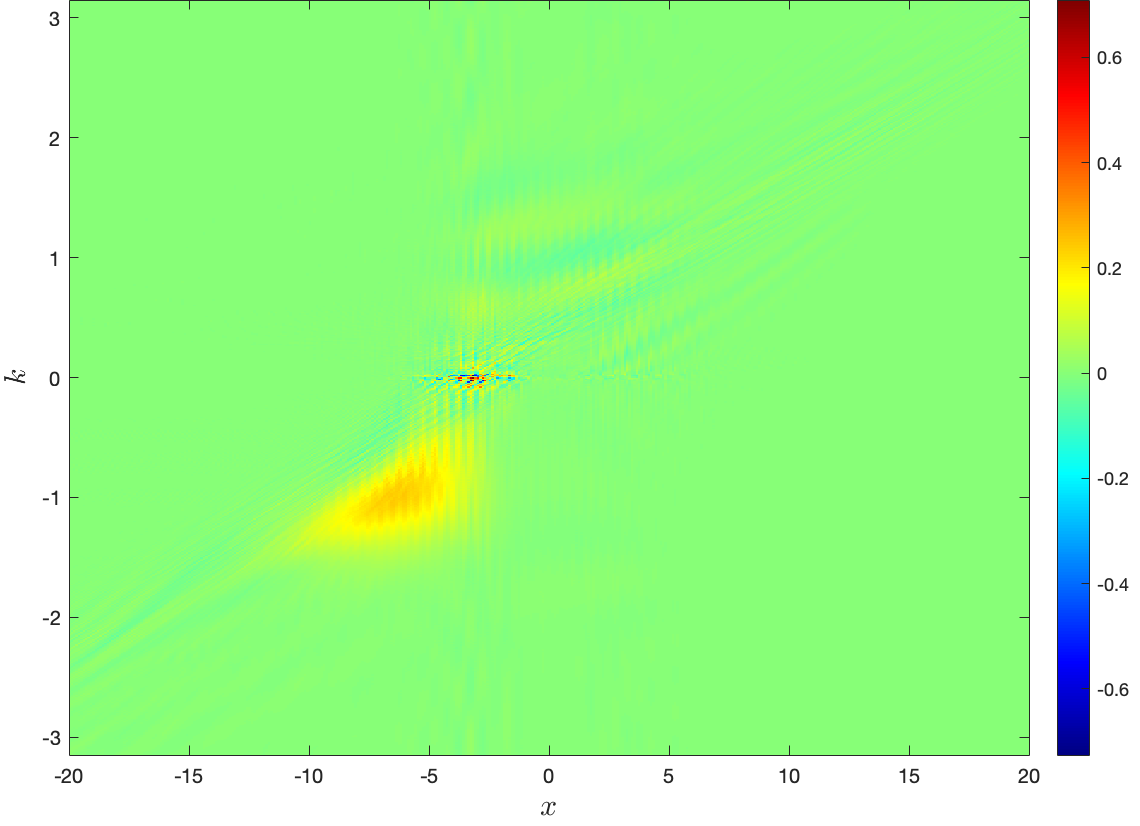}}\\
\caption[x-2]{\small The inverse square potential: The Wigner functions at different instants. 
The wave packet is almost totally reflected by the singular barrier.} 
\label{fig:fractional} 
\end{figure}

The Wigner kernel of the inverse square potential \eqref{V_fractional} is
\begin{equation}
\label{Vw_fractional}
V_w (x,k) = - \frac{4H }{\hbar} |k| \sin(2xk)
\end{equation} 
and plugging it into Eq.~\eqref{c}, we are able to get a close form for $c_\nu( x)$ like Eq.~\eqref{c_delta}. 
The parameters are: $[X_L, X_R] = [-30~\text{nm}, 30~\text{nm}]$, $[k_{min}, k_{max}] = [-\pi ~\text{nm}^{-1},\pi ~\text{nm}^{-1}]$, $N_k=512$, $Q=40$, $M=55$, $N_{um}=600$,  $\Delta t = 0.005~\text{fs}$, $H=1$, $x^0 =-5 ~\text{nm}$, $k^0=1 ~\text{nm}^{-1}$. Fig.~\ref{fig:fractional_error} verifies the spectral convergence against both $N_k$ and $M$. 
Fig.~\ref{fig:fractional} displays the quantum dynamics where the Gaussian wave packet is almost totally reflected after hitting the singular barrier.  The tunneling rate $P_r (t)$ are $0.01158$, $0.009115$, $0.006731$ and $0.002025$ at $t=2$, $4$, $5$, and $8$~fs, respectively, indicating transparently that it is difficult for the wave packet to pass through the barrier due to the strong singularity of the inverse power potential. This is very different from the scattering shown in Section~\ref{sec:power} with the inverse power potential which has much weaker singularity.
Moreover, severe oscillations of positive and negative Wigner functions clearly appear around the origin, which accords with the summary that the quantum behavior near the singularities is difficult to be measured.

\section{Conclusions}
\label{sec:conc}

With the help of the Wigner function approach for quantum mechanics --- a first-principle nonlocal model, we perform highly accurate numerical simulations of quantum dynamics under singular potentials, during which the nonlocal characteristic of Wigner function contributes to the attenuation of singularity of the potentials. Numerically converged Wigner functions under the Dirac delta function, the logarithmic, and the inverse power potentials are obtained with an operator splitting spectral method.
Many interesting quantum behaviors are also revealed during the scattering under these singular potentials. 
It should be noted that all existing Wigner simulations truncate the nonlocal integral in $k$-space, but 
the effect of such truncation on long-time simulations of quantum dynamics is hardly estimated in advance.
Instead, motivated by recently proposed adaptive technologies on unbounded domains \cite{xia2021frequency, doi:10.1137/20M1347711, chou2023adaptive}, we are developing numerical methods to solve the Wigner equation without truncating the nonlocal $k$-integral.

\section*{Acknowledgements}

This research was supported by the National Key R\&D Program of China (Nos. 2020AAA0105200, 2022YFA1005102) and the National Natural Science Foundation of China (Nos.~12288101, 11822102).
SS is partially supported by Beijing Academy of Artificial Intelligence (BAAI).

%
%



\end{document}